\documentclass[a4paper,fleqn]{cas-sc}

\usepackage[sort&compress,numbers]{natbib}

\def\tsc#1{\csdef{#1}{\textsc{\lowercase{#1}}\xspace}}
\tsc{WGM}
\tsc{QE}
\tsc{EP}
\tsc{PMS}
\tsc{BEC}
\tsc{DE}

\usepackage{siunitx} 
\usepackage[section]{placeins} 
\usepackage{subcaption} 
\usepackage{lineno}
\usepackage{setspace}

\graphicspath{{figs/}}

\newcommand{\tld}{$\sim$}

\begin{document}
\let\WriteBookmarks\relax
\def\floatpagepagefraction{1}
\def\textpagefraction{.001}

\shorttitle{In-situ Analysis of Deformation Dynamics in Cobalt}

\shortauthors{M. Knapek et~al.}

\title [mode = title]{In-situ Analysis of the Effect of Residual fcc Phase and Special Grain Boundaries on the Deformation Dynamics in Pure Cobalt}                      

\author[1]{Michal Knapek}[orcid=0000-0001-7079-2523]

\cormark[1]


\ead{michal.knapek@matfyz.cuni.cz}


\credit{Conceptualization, Formal Analysis, Investigation, Writing - original draft preparation, Supervision}

\author[1]{Peter Minárik}[]
\credit{Formal analysis, Investigation, Writing - review and editing}

\author[1]{Adam Greš}
\credit{Investigation, Writing - original draft preparation, Funding acquisition}

\author[1]{Patrik Dobroň}
\credit{Formal analysis, Investigation, Writing - review and editing}

\author[1]{Petr Harcuba}
\credit{Validation, Writing - review and editing, Funding acquisition}

\author[1]{Tomáš Tayari}
\credit{Investigation}

\author[1]{František Chmelík}
\credit{Conceptualization, Validation, Writing - review and editing, Supervision}

\affiliation[1]{organization={Charles University, Faculty of Mathematics and Physics},
 addressline={Ke~Karlovu~5}, 
 city={Prague},
 postcode={12116}, 
 country={Czech Republic}}

\cortext[cor1]{Corresponding author}


\begin{abstract}
Polycrystalline hcp metals -- an important class of engineering materials -- typically exhibit complex plasticity because of a limited number of slip systems. Among these metals, deformation is even more complicated in cobalt as it commonly contains residual fcc phase due to the incomplete martensitic fcc$\rightarrow$hcp transformation upon cooling. In this work, we employ a combination of in-situ (acoustic emission, AE) and ex-situ (scanning electron microscopy, SEM) techniques in order to examine deformation dynamics in pure polycrystalline cobalt varying in grain size and the content of residual fcc phase prepared using systematic thermal treatment and cycling. We reveal that the presence of the fcc phase and special \tld{}\ang{71} grain boundaries between different hcp martensite variants brings about higher deformability and strength. The fcc phase provides additional slip systems and also accommodates deformation via the stress-induced fcc$\rightarrow$hcp transformation during loading. On the other hand, special boundaries enhance structural integrity and suppress the formation of critical defects. Both these non-trivial effects can dominate over the influence of grain size, being a traditional microstructural variable. The ex-situ SEM experiments further reveal that the stress-induced fcc$\rightarrow$hcp transformation is sluggish and only partial even at high strains, and it does not give rise to detectable AE signals, unlike in other materials exhibiting martensitic transformation. In turn, these insights into cobalt plasticity provide new avenues for the microstructure and performance optimization towards the desired applications through the modern concept of grain boundary engineering.

\end{abstract}

\begin{graphicalabstract}
\includegraphics[width=\linewidth]{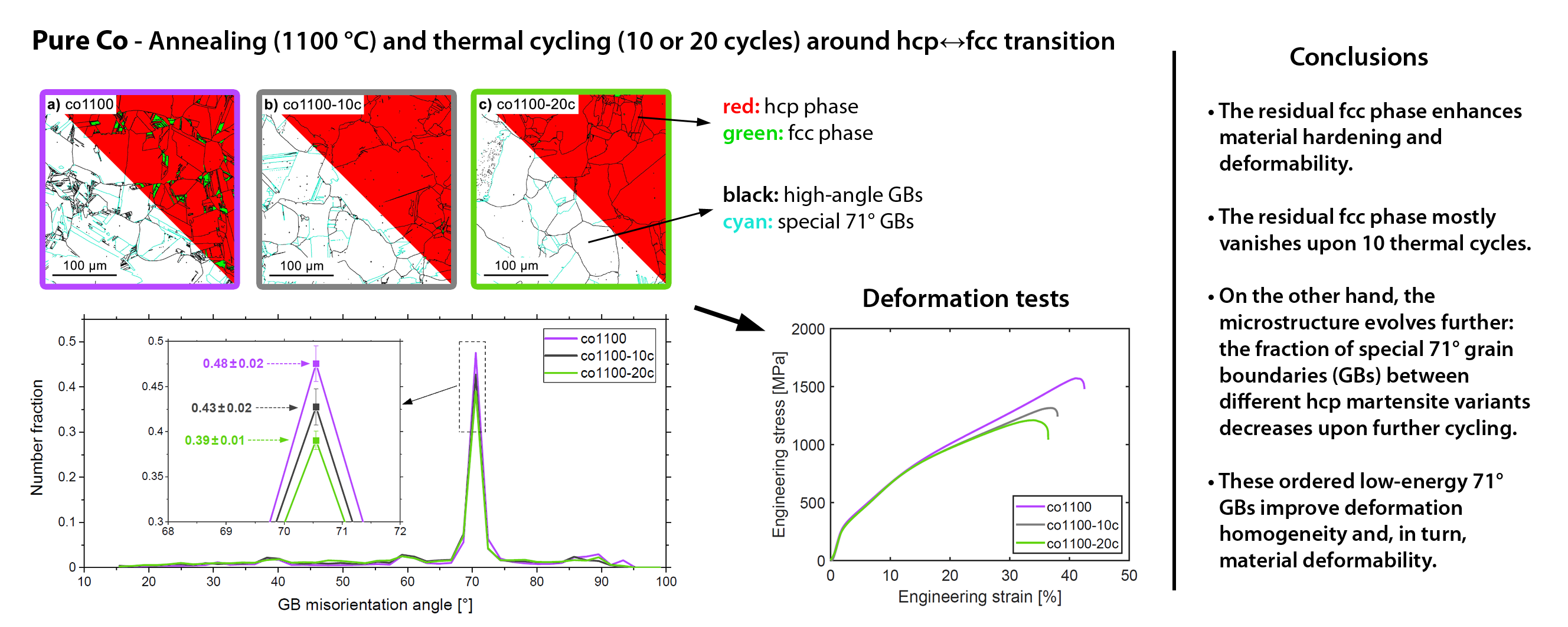}
\end{graphicalabstract}

\begin{highlights}
\item Residual fcc phase contributes to cobalt deformability and hardening.
\item Stress-induced fcc$\rightarrow$hcp transformation is slow and does not produce acoustic emission.
\item Not only amount but, especially, type of grain boundaries must be considered.
\item Thermal cycling eliminates fcc phase while special \ang{71} boundaries evolve further. 
\item Special \ang{71} boundaries (between hcp variants) enhance deformation homogeneity.
\end{highlights}


\maketitle


\section{Introduction}

Cobalt belongs to the group of metals with hexagonal close-packed (hcp) crystal lattice, which have been frequently utilized as engineering materials in numerous industries. Unlike other hcp metals, pure cobalt exhibits ferromagnetism and has a relatively high melting temperature \cite{betteridge_properties_1980}, and found its applications as an alloying element in magnetic alloys, high temperature superalloys, and carbide cutting tools \cite{sargent_deformation_1983}. Nowadays, cobalt is often utilized more directly for the production of batteries, in the information technology (computer memory and data storage), and also for the production of emerging class of various delicate miniaturized devices \cite{zheng_grain-size_2007, crone_brief_2008}. The design and reliability of (micro)mechanical parts essentially relies on the understanding of mechanical behavior involving the dynamics of operating deformation mechanisms in relation to the material microstructure.

The examination of mechanical behavior in polycrystalline cobalt is of high scientific interest due to the fact that cobalt exhibits an allotropic transformation between its low-temperature hcp and high-temperature face-centered cubic (fcc) crystal structures. This phase transformation belongs to the class of martensitic transformations \cite{edwards_xray_1943,troiano_transformation_1948}, i.e. it is a first-order diffusionless crystallographic transformation that is (in ideal case) reversible, and its existence might render the resulting material microstructure peculiarly complex. The transformation is realized by a motion of interfaces between two distinct phases as the atoms in the parent lattice rearrange into an energetically more favorable structure upon heating or cooling across the transformation temperature. The collective atomic shift during this transformation brings about shape and symmetry changes of the crystal following the Shoji-Nishijama geometric relation: $\{0001\}_\textrm{hcp}\parallel\{111\}_\textrm{fcc}$ and $\langle11\bar{2}0\rangle_\textrm{hcp}\parallel\langle110\rangle_\textrm{fcc}$, providing four possible variants of the hcp martensite corresponding to four $\{111\}_\textrm{fcc}$ planes \cite{nishiyama_crystallography_1978,laughlin_physical_2014}. Martensitic transformations have long been observed in several other metals and metallic alloys, such as steels, Ti, Ni-Ti alloys, and complex concentrated alloys, and were commonly studied using electron microscopy techniques \cite{schryvers_electron_1997,xie_transmission_2005,chang_high-temperature_2019}.

In cobalt, the allotropic martensitic hcp$\leftrightarrow$fcc transformation occurs at a relatively low temperature around \SI{420}{\celsius}. This temperature is only an approximation by roughly averaging the reported values of martensite start and austenite start temperatures, as they can exhibit a thermal hysteresis of as much as \SI{40}{\celsius} in polycrystalline cobalt, based on its purity, microstructural features, and thermo-mechanical history \cite{bauer_kinetics_2011, song_recrystallization_2022}. Such a large hysteresis results from quite a low enthalpy change upon transformation (\tld\SI{500}{\joule\per\mol} on heating, \tld\SI{370}{\joule\per\mol} on cooling) that also accounts for its sluggishness \cite{betteridge_properties_1980, toledano_theory_2001}. Most importantly, upon cooling across the martensite start temperature, a significant fraction of the high-temperature fcc phase can remain in the microstructure (in other words, the fcc$\rightarrow$hcp transformation is incomplete). On the other hand, it was reported that the fraction of the retained fcc phase in cobalt can be reduced by repeated thermal cycling across the transition temperatures, thus offering additional means of microstructure tailoring \cite{bauer_kinetics_2011, munier_evolution_1990,bidaux_study_1987,kuang_latent_2000}. 

Metals with hcp structure typically show quite complex deformation dynamics owing to only a limited number of slip systems in comparison with metals with cubic structures \cite{mises_mechanik_1928}. Naturally, with the presence of additional phase, the deformation processes become even more complicated in cobalt. This is particularly because there is another crucial point to consider: the residual fcc phase is unstable and during loading it transforms to the stable room temperature hcp modification, thus making the deformation dynamics of cobalt even more perplexing \cite{sanderson_deformation_1972, dubos_temperature_2020}. Consequently, the residual fcc phase acts as another intriguing microstructural parameter affecting the mechanical performance of the material in addition to typical parameters such as grain size, crystallographic texture, or residual stress. 

Targeted studies on deformation dynamics in relation to the microstructure of polycrystalline cobalt are, however, rather rare. Most studies were conducted several decades ago when the technology of microstructural observations was still rather undeveloped \cite{holt_high_1968,sanderson_deformation_1972}. Some authors focused on cobalt single crystals \cite{holt_high_1968, holt_influence_1972, korner_weak-beam_1983}, thin films \cite{marx_strain-induced_2016}, and nano- or microcrystalline samples \cite{sort_microstructural_2003, edalati_high-pressure_2013, barry_microstructure_2014}; or, lately, they studied dynamic recrystallization at elevated temperatures \cite{kapoor_aspects_2009,song_recrystallization_2022}. There are only several works investigating directly the activity of deformation mechanisms including mechanical twinning, primarily in tension \cite{seeger_plastische_1963,zhang_deformation_2010,fleurier_size_2015, martinez_tem_2017}.

In our recent work \cite{knapek_effect_2020} we systematically studied the microstructure variations in polycrystalline cobalt as a function of applied thermal treatments (isothermal annealing at different temperatures and thermal cycling around the hcp$\leftrightarrow$fcc transformation) using scanning electron microscopy (SEM) and thermal analyses. We also examined the basic microstructure-compressive behavior relations that suggested a non-trivial dependence of deformation dynamics on the fraction of the residual fcc phase. In order to gain deeper understanding of these effects, we apply in this study a combination of advanced experimental techniques including in-situ acoustic emission (AE) and ex-situ electron backscatter diffraction (EBSD) recorded during loading. In correlation with deformation curves, these original data sets allowed us to examine the effect of the fcc phase and related microstructural features on the dynamics of operating deformation mechanisms and their evolution with increasing strain.  

\section{Materials and Methods}

\begin{figure}[]
\centering
    \begin{subfigure}[b]{0.4\textwidth}
        \centering
        \includegraphics[width=\linewidth]{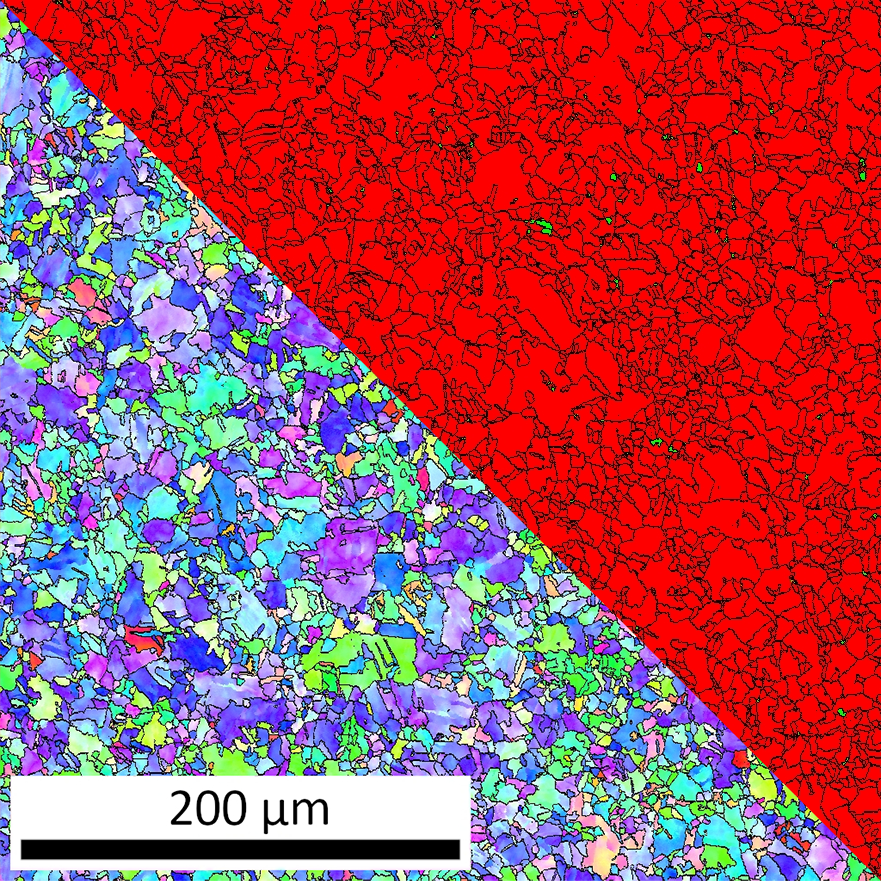}
    \end{subfigure} \hspace{1pt}
    \begin{subfigure}[b]{0.3\textwidth}
        \centering
        \raisebox{0.2\height}{\includegraphics[width=\linewidth]{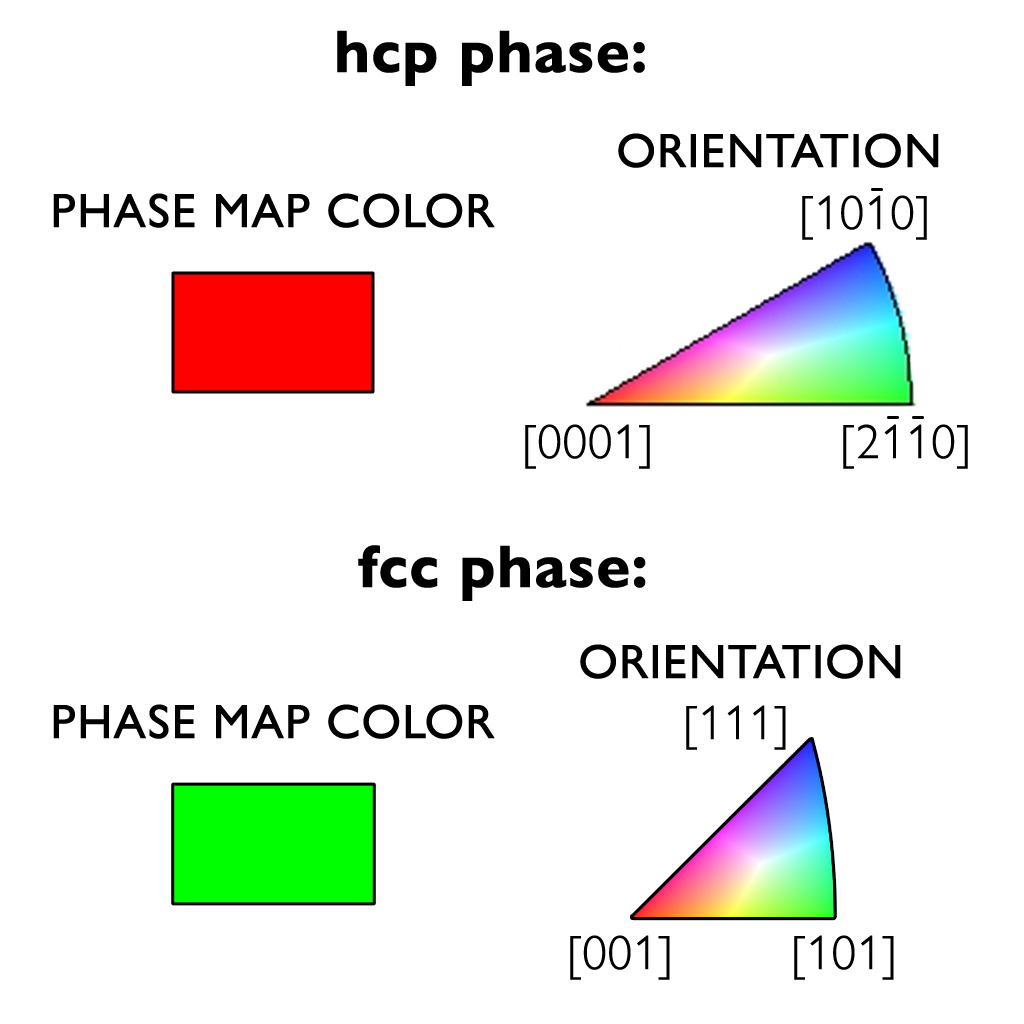}}
    \end{subfigure}
    \caption{The EBSD orientation map of the initial microstructure of the as-drawn polycrystalline cobalt. The micrograph is diagonally split to show both the EBSD orientation maps (left bottom) and the EBSD phase maps (top right) of the studied area. The map plane is perpendicular to the drawing direction.}
    \label{EBSD_init}
\end{figure}

The samples for the analyses were prepared from as-drawn pure cobalt rods (\SI{200}{mm} in length, \SI{6.35}{mm} in diameter) purchased from Goodfellow Cambridge Ltd. (Huntingdon, England). The material purity was \SI{99.9}{\percent}, with the following quoted impurities (in ppm): Fe -- 180, Ni -- 800, C -- 30, S -- 150. The initial as-drawn microstructure (before any thermal processing) is presented in Fig. \ref{EBSD_init} by means of EBSD orientation map and phase map perpendicular to material (i.e. rod) drawing direction. The material contained residual stresses from the drawing procedure, the grain size was \tld$\SI{9}{\um}$, and it can be inferred from the preferential grain orientation that it featured a common texture of the drawn or extruded hcp materials \cite{dobron_grain_2011} with basal planes oriented preferentially along the processing direction. The texture magnitude was \tld{}3 multiples of random (the inverse pole figures were published elsewhere \cite{knapek_effect_2020}). Also, in this as-received condition there was almost no residual fcc phase in the material. 

Sets of samples with a height of \SI{9}{mm} were cut from the rods and were annealed under vacuum in the vertical furnace (Nabertherm RHTV 120-600, Lilienthal, Germany) for \SI{1}{h} and water-quenched. The annealing temperatures were \SI{600}{\celsius} up to \SI{1100}{\celsius} with \SI{100}{\celsius} step. A part of the  samples annealed at each temperature were further subjected to thermal cycling around the cobalt hcp$\leftrightarrow$fcc phase transformation temperature (\tld\SI{420}{\celsius}) in order to vary and, to some extent, stabilize the microstructure in terms of the grain structure and the fcc phase fraction. The thermal cycling was performed using the Linseis L75 PT vertical thermodilatometer (Selb, Germany), under an Ar atmosphere. One thermal cycle consisted of heating from \SI{300}{\celsius} to \SI{550}{\celsius} and cooling back to \SI{300}{\celsius} at the rate of \SI{5}{\celsius\per\minute}. The samples were subjected to either 10 or 20 such cycles. Henceforth, the ``coXXXX-YYc'' naming scheme will be used for the samples, where ``XXXX'' is replaced with the sample annealing temperature and ``YY'' is replaced with the number of cycles. Samples not subjected to thermal cycling will be labelled as ``coXXXX''.

The microstructure of the prepared materials was investigated using SEM (ZEISS Auriga Compact FIB-SEM, Jena, Germany), primarily in terms of EBSD (EDAX EBSD Velocity camera, Berwyn, USA) with sufficient resolution for the visual examination and, especially, for the calculation of various quantities of interest (grain size, twin fraction, fcc phase fraction, misorientation angles) up to scan areas of 2000$\times$\SI{1000}{\um^2} with step size of \SI{2}{\um}. Before the SEM observations, the sample surfaces were prepared by grinding using grit emery papers and polishing using diamond suspensions with decreasing particle sizes down to \SI{1}{\micro\meter}. The final step was ion polishing performed using Leica EM RES102 (Leica Mikrosysteme, Wetzlar, Germany). The EBSD data were cleaned and processed using the EDAX OIM Analysis 8 software. The procedure consisted of confidence index (CI) standardization, phase neighbor correlation, and one iteration of grain dilatation. For further analysis only points with CI of at least 0.1 were considered. Misorientations higher than \SI{15}{\degree} were considered a grain boundary \cite{lejcek_grain_2010} (i.e. including possible twin and phase boundaries) and the average grain size was weighted by the area fraction of the grains. The area fraction of grain types (twins/phases) and the number fraction of type of grain boundaries were calculated using the EDAX OIM Analysis 8 software in a standard way, i.e. as a fraction of map points respective to all the points identified as belonging to any grain or all the points identified as grain boundaries, respectively. 

The compression tests were performed using the Instron 5882  universal testing machine (Northwood, Massachusetts, USA) at room temperature and with initial strain rate of $\dot{\varepsilon} = 10^{-3}$\ \si{\per\second}. To prevent sample barrelling, Apiezon~M lubricant (M\&I Materials Ltd, Manchester, UK) was applied to deformation platens in the sample contact areas. Several tests were performed to verify the repeatability of the results. The machine stiffness was subtracted from measured data. The systematic error was determined to be \tld{}\SI{3}{\percent}. During deformation, AE data were also acquired using the PCI-2 board, the 2/4/6 switch-selectable preamplifier, and the PICO miniature broadband sensor (all from Physical Acoustics Corporation, New Jersey, USA). The AE sensor lubricated with Apiezon-M grease was attached to the deformation platen in the vicinity of the sample using a plastic clip, securing equal conditions for all the tests. Event detection and parametrization was performed on the recorded raw AE signals using the standard threshold-based procedure to suppress the influence of background noise. 

Additional campaign of interrupted (``ex-situ'') compression tests combined with EBSD was performed in order to analyze the dynamics of operating deformation mechanisms in relation to the samples' microstructure. The samples were prepared following the same procedure of preparation for EBSD as described above. The EBSD measurement was first performed on the initial undeformed state of each sample. Subsequently, the sample was placed in the testing machine, deformed to certain strain, and the same area was analysed by EBSD again. This process was repeated for the strains of \SI{1}{\percent}, \SI{4}{\percent}, \SI{10}{\percent}, \SI{20}{\percent}, and \SI{30}{\percent}. The EBSD data were further cleaned and processed using the EDAX OIM Analysis 8 software. The data processing consisted of one iteration of Neighbor pattern averaging \& reindexing (NPAR), reindexing the points with CI lower than 0.1, and the clean-up steps described above. 


\section{Results}

\begin{figure}[b!]
\centering
    \begin{subfigure}[b]{0.5\textwidth}
        \centering
        \includegraphics[width=\linewidth]{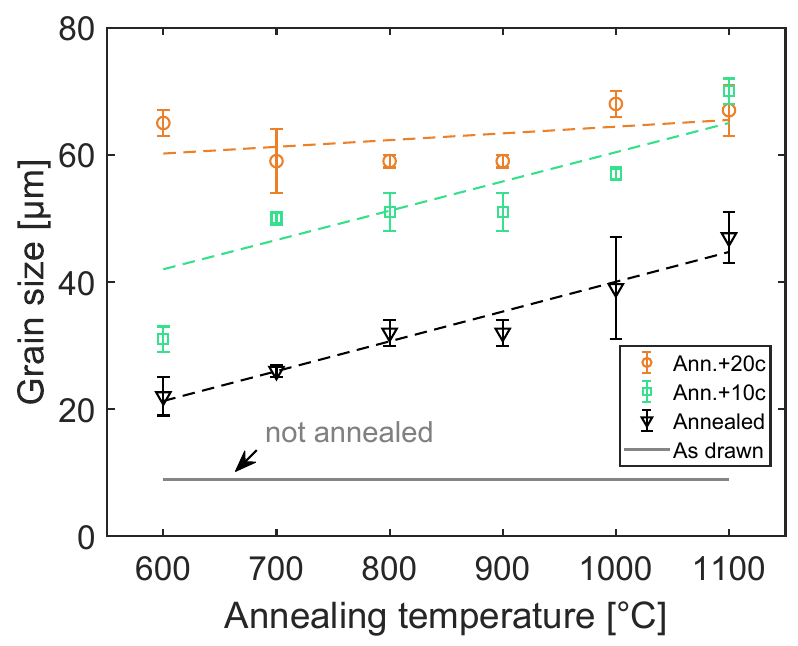}
    \end{subfigure} \hspace{1pt}
    \caption{Grain sizes of the cobalt samples subjected to different thermal treatments: annealing at \SI{600}{\celsius}--\SI{1100}{\celsius} and annealing at \SI{600}{\celsius}--\SI{1100}{\celsius} followed by thermal cycling around the hcp$\leftrightarrow$fcc phase transformation temperature (10 or 20 cycles).}
    \label{GS}
\end{figure}

After the thermal treatment and cycling, the microstructural features of the samples were examined in detail. Fig. \ref{GS} shows the grain sizes (defined by boundaries with misorientation of \SI{>15}{\degree} as well as fcc/hcp boundaries) of all the prepared material conditions, namely the annealed co600--co1100 samples and the annealed+thermally cycled samples co600-10c--co1100-10c and co600-20c--co1100-20c. It can be deduced that annealing led to a gradual increase in the grain size as a function of rising annealing temperature from \tld$\SI{9}{\um}$ (as-drawn) to \tld$\SI{20}{\um}$ (co600) up to \tld$\SI{45}{\um}$ (co1100). Upon 10 thermal cycles, the grain size increased further to \tld$\SI{30}{\um}$ for co600-10c up to \tld$\SI{65}{\um}$ for co1100-10c. Finally, after 20 cycles, the grain size became rather uniform across the materials co600-20c--co1100-20c and reached the value of \tld$\SI{65}{\um}$. It should be pointed out that, interestingly, the grain size increased quite significantly upon cycling (up to an apparent saturation after 20 cycles) even though the temperature of prior isothermal annealing was much higher. It was, however, reported (e.g. \cite{sanderson_deformation_1972, betteridge_properties_1980, kapoor_aspects_2009,song_recrystallization_2022}) that the recrystallization of cobalt can take place at the temperature as low as of \tld{}\SI{350}{\celsius} and the prolonged thermal cycling at mild temperature thus appears to dominate over a relatively short $\SI{1}{\hour}$ annealing at up to \SI{1100}{\celsius}. In addition, as elaborated already in the Introduction, it was asserted that annealing and thermal cycling significantly affect the amount of residual fcc phase. Our recent research \cite{knapek_effect_2020} showed that while the annealed materials co600--co1100 contained a considerable amount of fcc phase up to \tld{}\SI{10}{\percent} all the thermally cycled samples featured less than \SI{1}{\percent} fcc after 10 cycles and less than \SI{0.5}{\percent} fcc after 20 cycles (for details and all the EBSD maps see \cite{knapek_effect_2020,gres_experimental_2023}). 

\begin{figure}[b!]
\centering
    \begin{subfigure}[b]{0.25\textwidth}
        \centering
        \includegraphics[width=\linewidth]{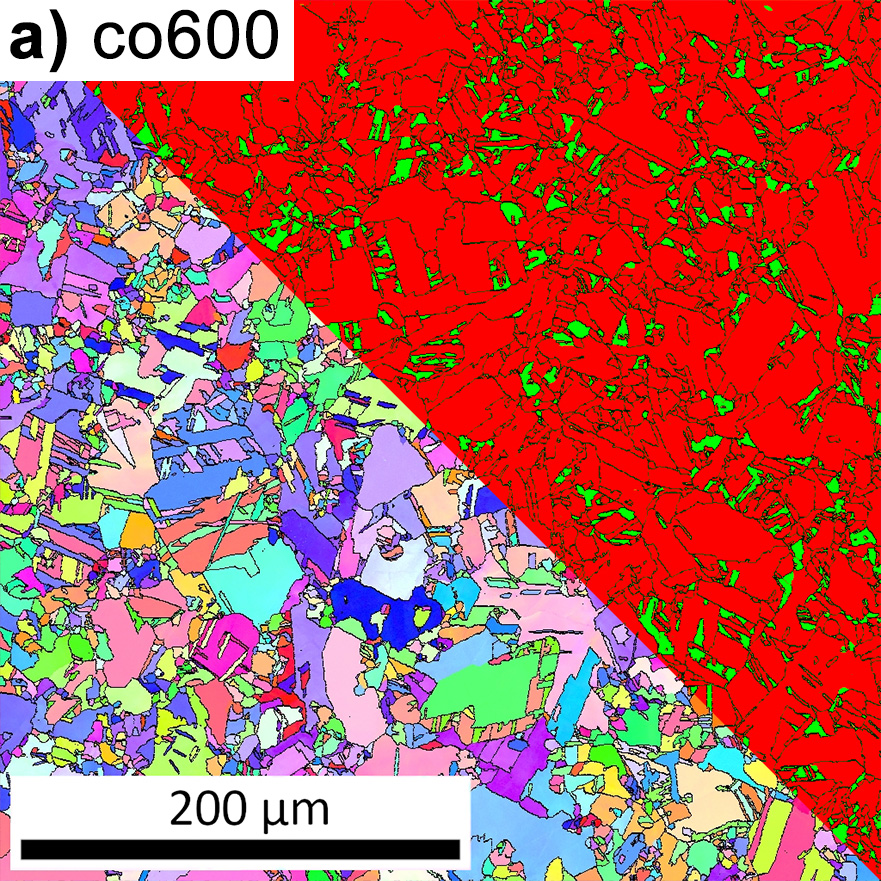}
    \end{subfigure} \hspace{1pt}
    \begin{subfigure}[b]{0.25\textwidth}
        \centering
        \includegraphics[width=\linewidth]{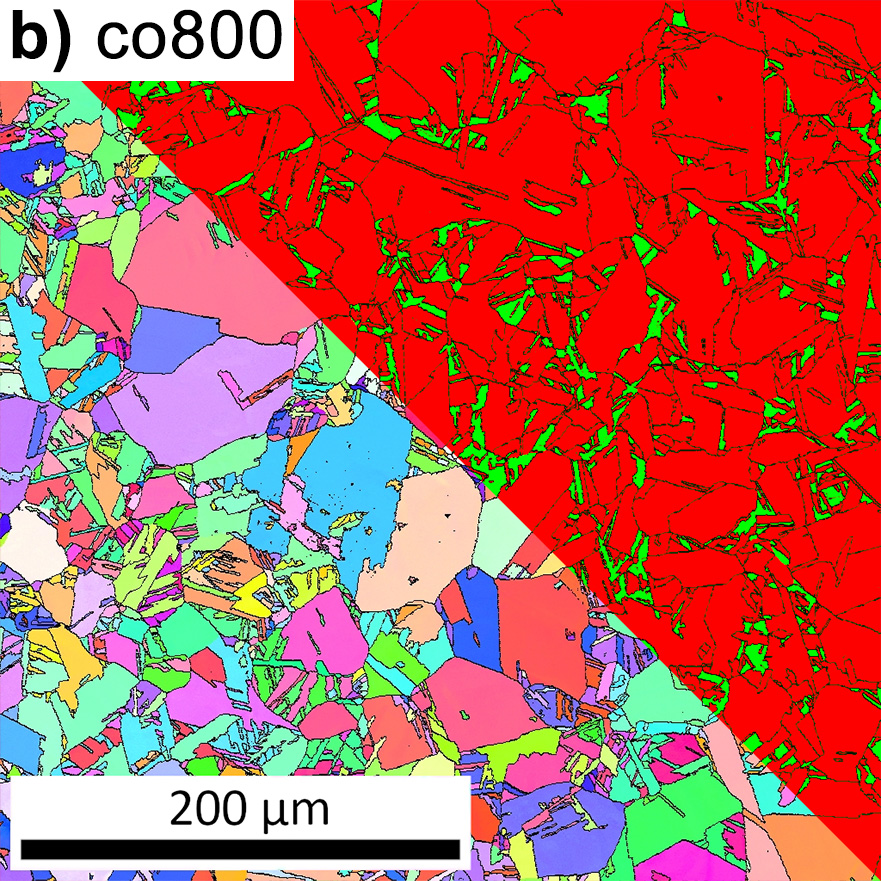}
    \end{subfigure}

    \vspace{0.2cm}

    \begin{subfigure}[b]{0.25\textwidth}
        \centering
        \includegraphics[width=\linewidth]{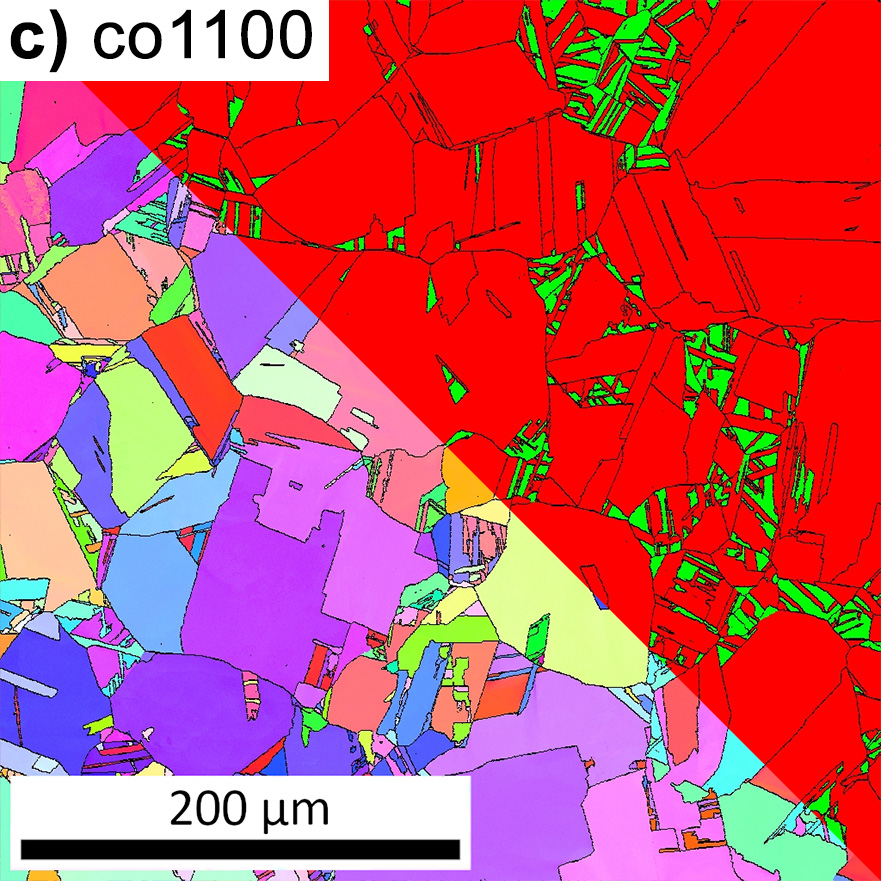}
    \end{subfigure} \hspace{1pt}
    \begin{subfigure}[b]{0.25\textwidth}
        \centering
        \includegraphics[width=\linewidth]{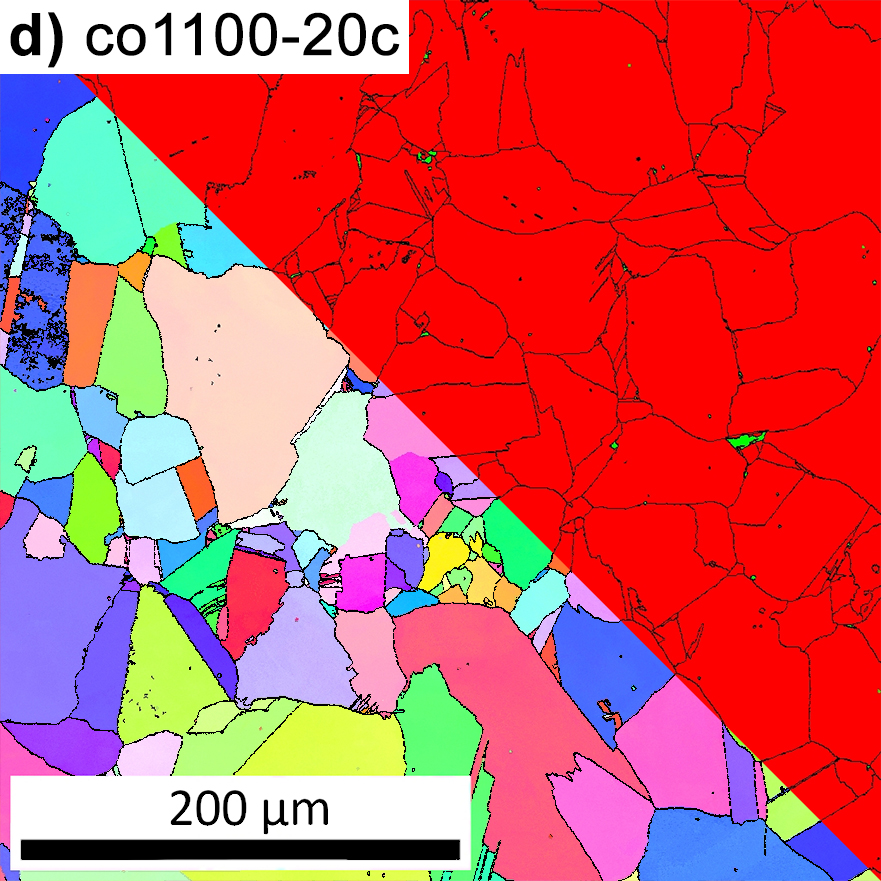}
    \end{subfigure}

    \vspace{0.2cm}

    \begin{subfigure}[b]{0.52\textwidth}
        \centering
        \includegraphics[width=\linewidth]{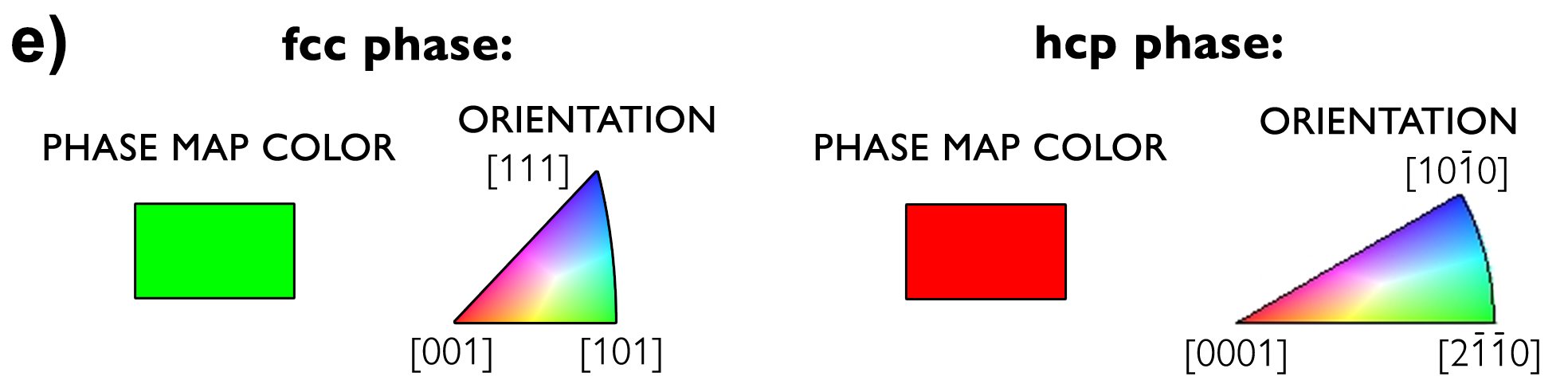}
    \end{subfigure}

    \caption{Microstructure of the selected thermally-treated samples of pure polycrystalline cobalt. The micrographs are diagonally split to show both the EBSD orientation maps (left bottom) and the EBSD phase maps (top right) of the studied areas. The materials a) co600, b) co800, and c) co1100 contain similar fraction of the fcc phase and vary in the grain size, whereas the samples c) co1100 and d) 1100-20c feature a similar grain size but differ in the fcc content, thus allowing for the examination of both effects.}
    \label{EBSD_TT}
\end{figure}

Based on these considerations, four representative materials, namely co600, co800, co1100, and co1100-20c, were selected for the detailed examination of deformation dynamics in terms of in-situ and ex-situ testing. The microstructures of the four selected materials are displayed in Fig. \ref{EBSD_TT} showing the EBSD orientation maps (left bottom part of the diagonally split images) and phase maps (top right part). Using this selected set of samples, we could examine (i) the effect of grain size on the deformation dynamics while the fcc phase content in the samples is rather similar ---  co600 (fcc \tld{}\SI{6}{\percent}, grain size \tld$\SI{20}{\um}$), co800 (fcc \tld{}\SI{8}{\percent}, grain size \tld$\SI{30}{\um}$), and co1100 (fcc \tld{}\SI{6}{\percent}, grain size \tld$\SI{50}{\um}$), and (ii) the effect of the fcc phase in the samples with similar grain size --- co1100 (fcc \tld{}\SI{6}{\percent}, grain size \tld$\SI{50}{\um}$) and co1100-20c (fcc \SI{<0.5}{\percent}, grain size \tld$\SI{65}{\um}$). The results seemingly indicate that upon 20 thermal cycles a rather stabilized microstructure is produced, independently of the previous thermal treatment. Apart from the above-discussed microstructural features, the micrographs in Fig. \ref{EBSD_TT} show randomization of the crystalographic texture (cf. Fig. \ref{EBSD_init}) due to thermal treatment, leading to only a very weak texture which was determined to be of <1.5 multiples of random for all the thermally treated materials, Fig. \ref{EBSD_TT} (respective inverse pole figures were published in our previous studies \cite{knapek_effect_2020,gres_experimental_2023}).   

\begin{figure}[]
\centering
    \begin{subfigure}[b]{0.5\textwidth}
        \centering
        \includegraphics[width=\linewidth]{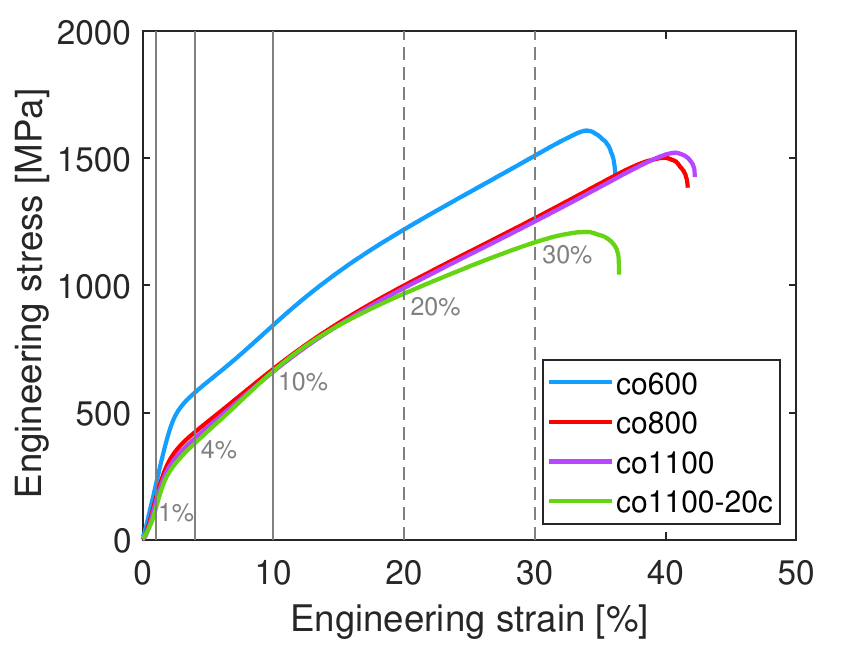}
    \end{subfigure} \hspace{1pt}
    \caption{Compression curves of the selected materials co600, co800, co1100, and co1100-20c. Grey vertical lines mark the points for supplementary ex-situ EBSD tests, as will be elaborated below.}
    \label{def}
\end{figure}

\begin{figure}[]
\centering
    \begin{subfigure}[b]{0.45\textwidth}
        \centering
        \includegraphics[width=\linewidth]{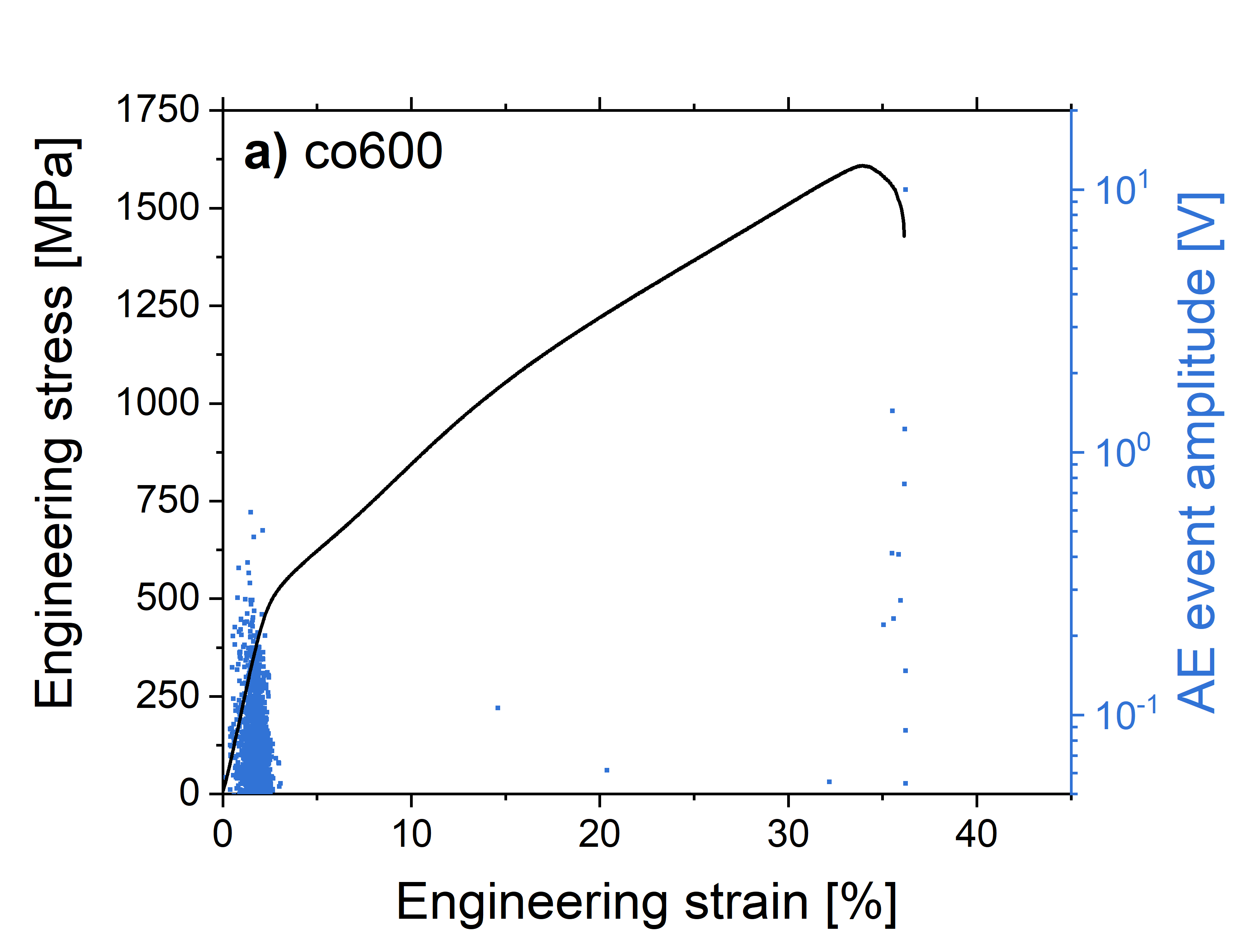}
    \end{subfigure} \hspace{1pt}
    \begin{subfigure}[b]{0.45\textwidth}
        \centering
        \includegraphics[width=\linewidth]{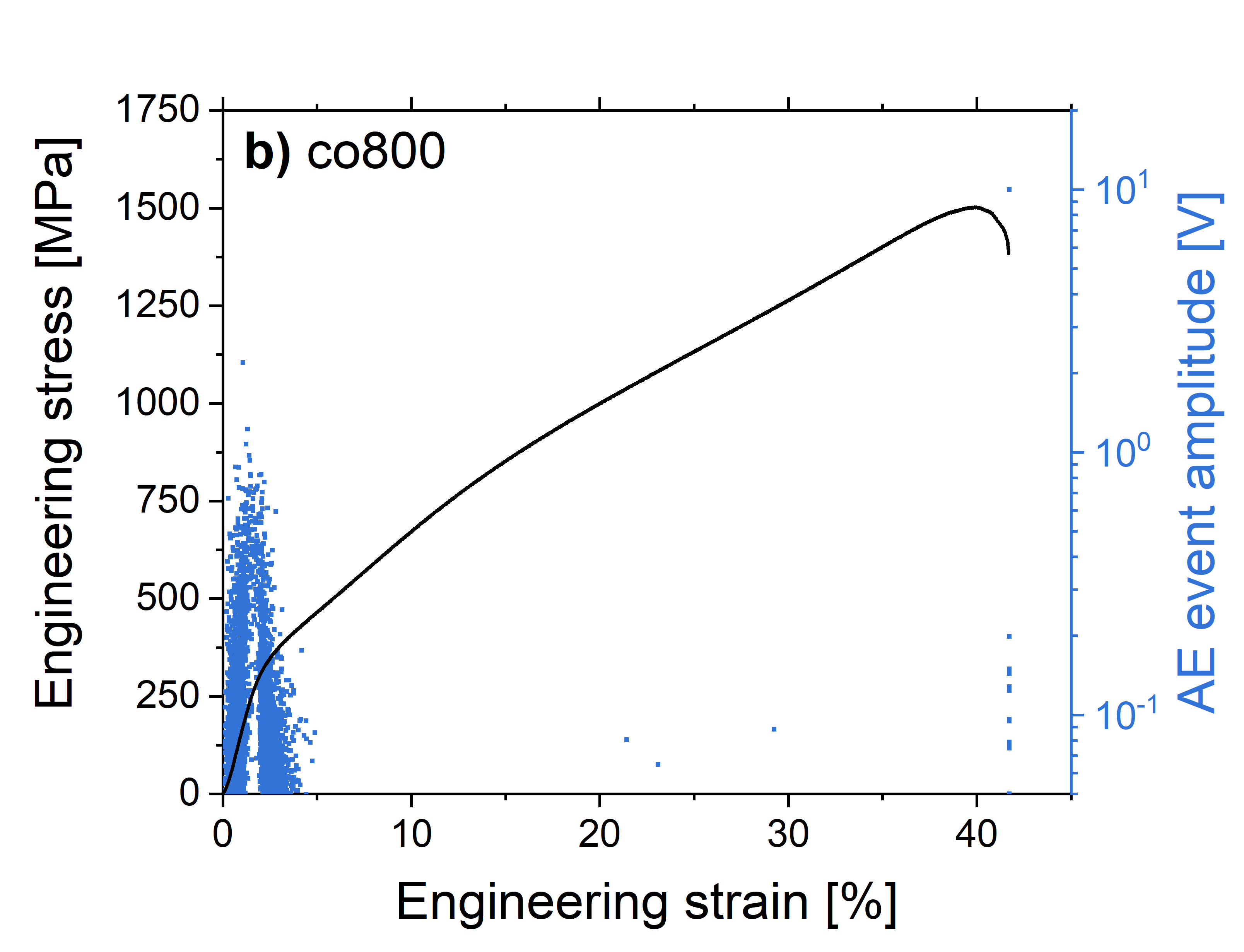}
    \end{subfigure}

    \vspace{0.05cm}

    \begin{subfigure}[b]{0.45\textwidth}
        \centering
        \includegraphics[width=\linewidth]{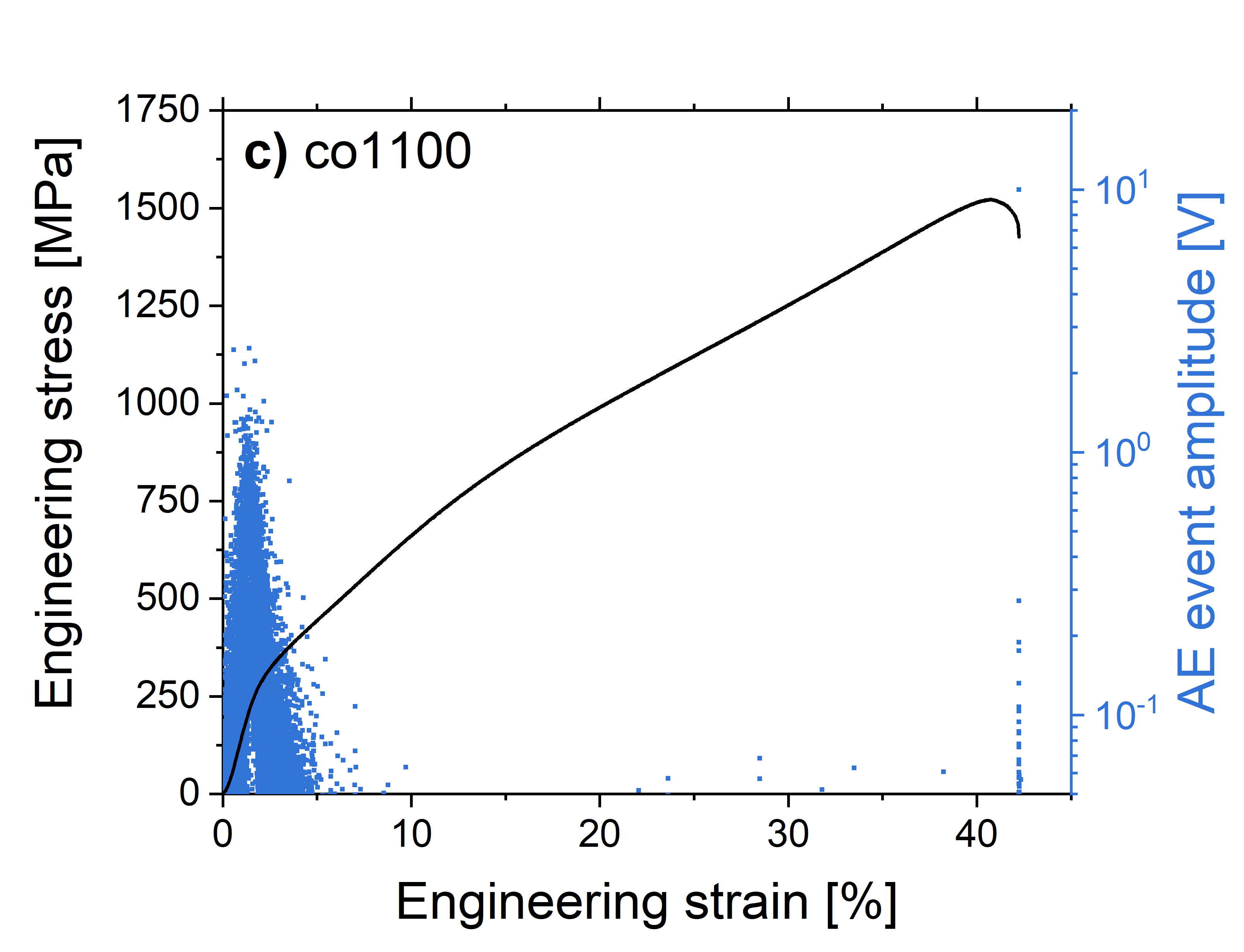}
    \end{subfigure} \hspace{1pt}
    \begin{subfigure}[b]{0.45\textwidth}
        \centering
        \includegraphics[width=\linewidth]{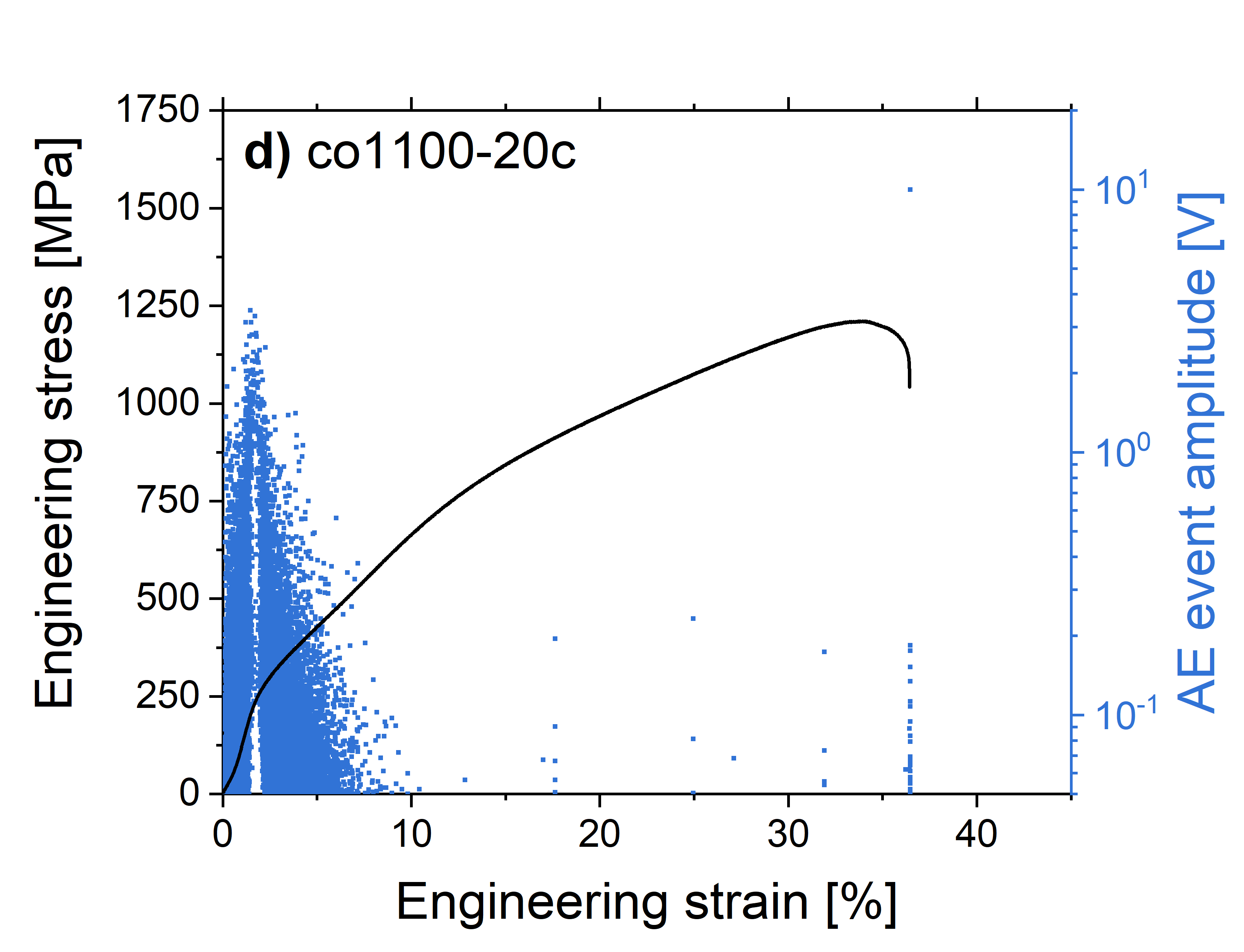}
    \end{subfigure}
    
    \caption{The AE response recorded during compression loading of samples a) co600, b) co800, c) co1100, and d) co1100-20c.}
    \label{AE}
\end{figure}

The representative engineering deformation curves resulting from the compression tests are presented in Fig.~\ref{def}. The co600 material exhibited the highest flow stress and, accordingly, the greatest yield stress ($\sigma_{0.2}$\tld$\SI{550}{MPa}$) and ultimate compressive stress ($\sigma_{\mathrm{max}}$\tld$\SI{1610}{MPa}$), likely owing to the fact that this material retained a rather fine microstructure due to lower annealing temperature. The deformability of co600 reached $\varepsilon_{\mathrm{max}}$\tld{}\SI{34}{\percent}. On the other hand, both the co800 and co1100 materials deformed similarly and exhibited $\sigma_{0.2}$ of \tld$\SI{380}{MPa}$ and \tld$\SI{350}{MPa}$, and $\sigma_{\mathrm{max}}$ of \tld$\SI{1500}{MPa}$ and \tld$\SI{1520}{MPa}$, respectively. $\varepsilon_{\mathrm{max}}$ of co800 was \tld{}\SI{40}{\percent} and it the case of co1100 it was \tld{}\SI{41}{\percent}. Increasing annealing temperature thus resulted in only slight lowering of $\sigma_{0.2}$, increase in $\sigma_{\mathrm{max}}$, and similar $\varepsilon_{\mathrm{max}}$, approaching the experimental error. Finally, the thermally cycled material co1100-20c had slightly inferior $\sigma_{0.2}$ of \tld$\SI{320}{MPa}$ but considerably lower $\sigma_{\mathrm{max}}$ and $\varepsilon_{\mathrm{max}}$ of \tld$\SI{1210}{MPa}$ and \tld{}\SI{34}{\percent}, respectively. Given the most extensive thermal treatment and resulting microstructure stabilization in this material, the inferior deformability might be seemingly unanticipated. The origin of this observation will be examined in detail below via ex-situ EBSD during deformation tests interrupted at points signified by vertical lines in Fig. \ref{def} as well as via ancillary examination of the ``intermediate'' material co1100-10c. 

All the compression tests were accompanied by the acquisition of AE signals originating from the dynamic microstructural changes in the loaded materials. Fig. \ref{AE} shows the deformation curves together with the evolution of amplitude of AE events recorded simultaneously during loading of the studied materials (note the logarithmic scale). Practically all the AE activity takes place in the early stages of deformation around the yield point. It is, however, worth noting that the amplitudes are already high during the seemingly elastic stage of deformation. After the yield point, the activity diminishes quite rapidly and beyond this active interval only a few sporadic events appear until the end of the test when some cracking-induced AE is observed due to the development of critical crack leading to failure. The amplitudes of events as well as the time windows of AE activity observed in the vicinity of yielding increase with higher annealing temperature from the samples co600 through co800 to co1100, and are the greatest in the case of thermally cycled material co1100-20c. The AE technique is particularly sensitive to rapid plastic deformation events (twin nucleation, dislocation avalanches, etc.) whose operation is, in turn, related to the microstructural features of the material. These in-situ data can thus be best interpreted in the context of evolving deformation mechanisms.

\begin{figure}[]
\centering
    \begin{subfigure}[b]{0.25\textwidth}
        \centering
        \includegraphics[width=\linewidth]{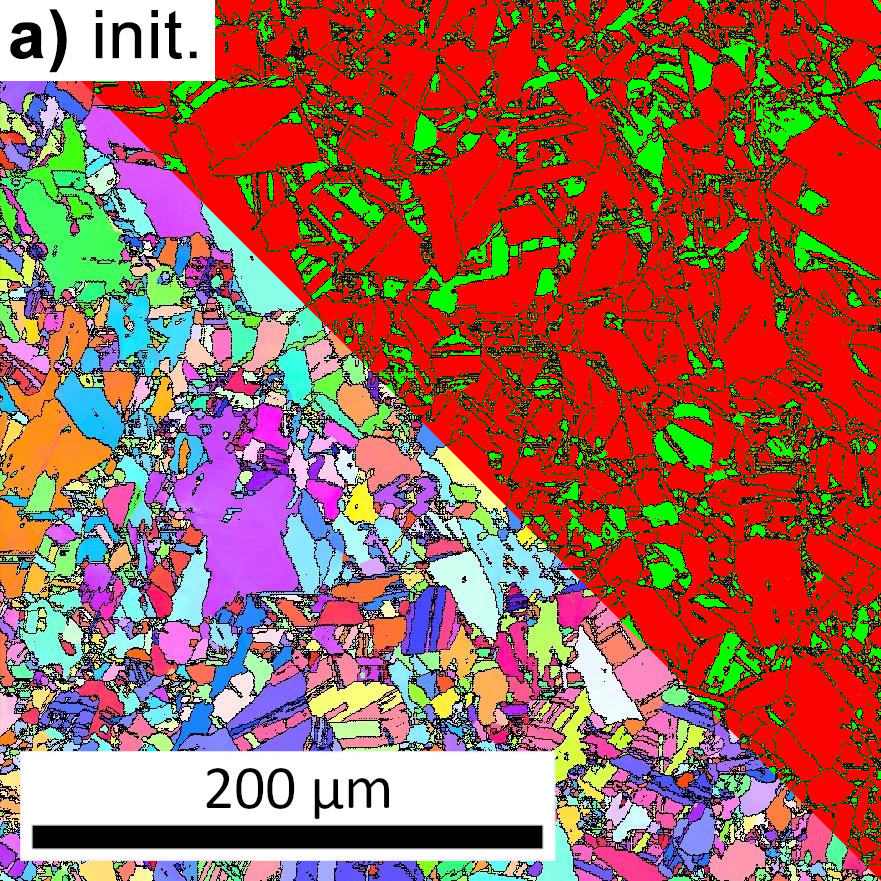}
    \end{subfigure} \hspace{1pt}
    \begin{subfigure}[b]{0.25\textwidth}
        \centering
        \includegraphics[width=\linewidth]{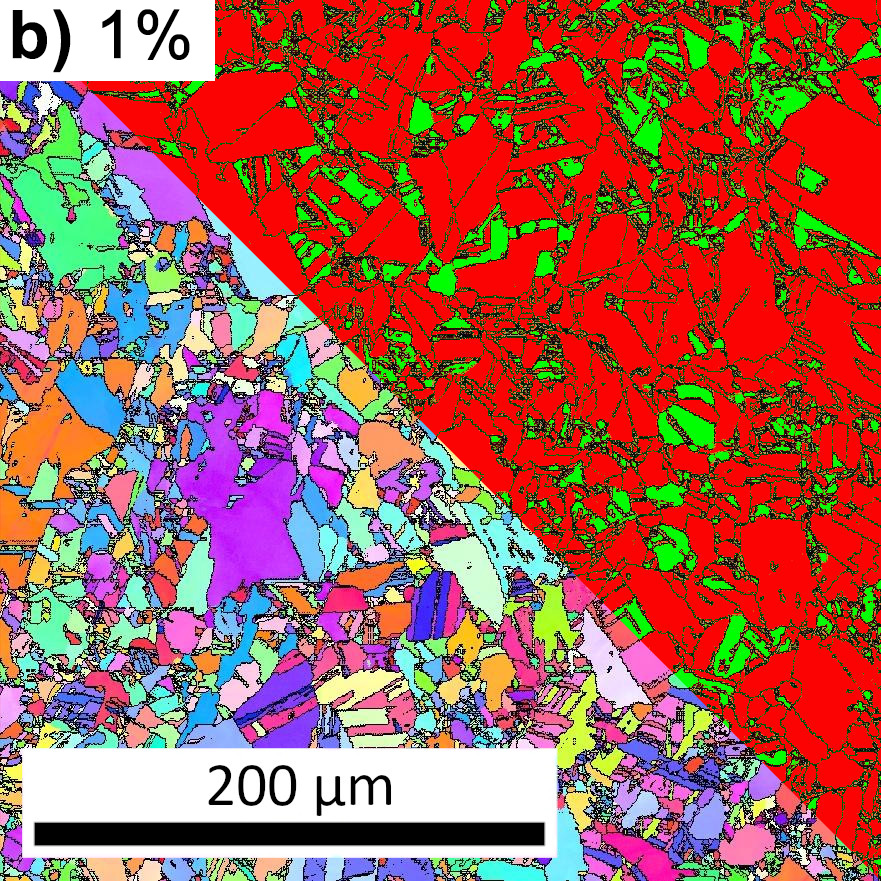}
        
    \end{subfigure}

    \vspace{0.2cm}

    \begin{subfigure}[b]{0.25\textwidth}
        \centering
        \includegraphics[width=\linewidth]{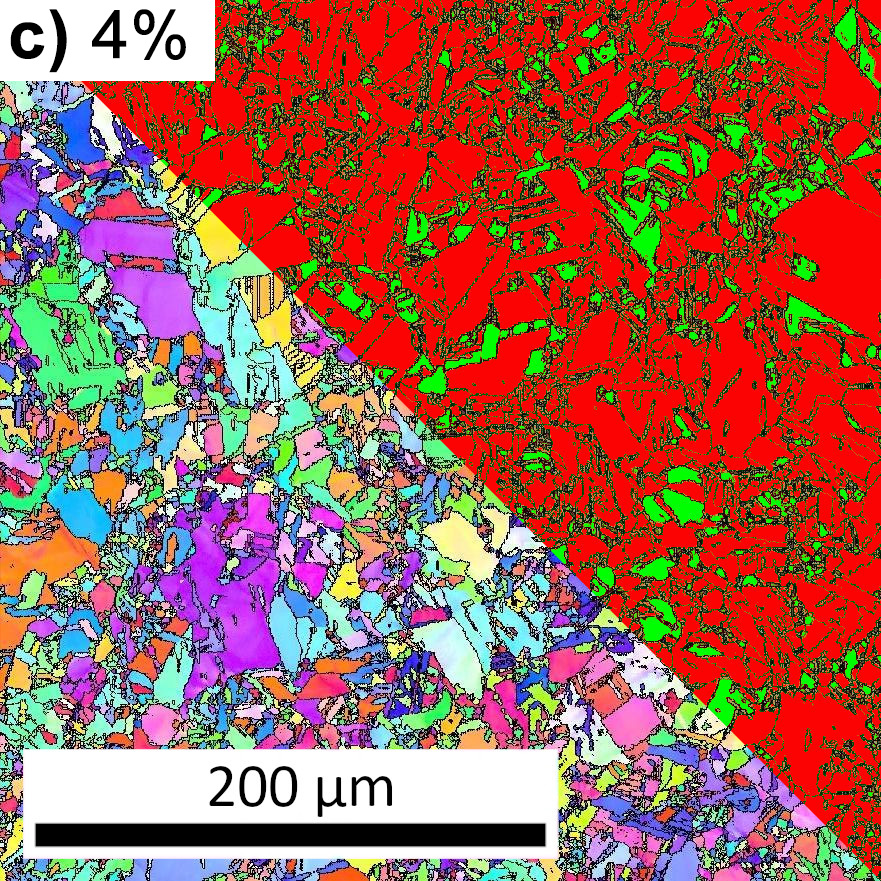}
    \end{subfigure} \hspace{1pt}
    \begin{subfigure}[b]{0.25\textwidth}
        \centering
        \includegraphics[width=\linewidth]{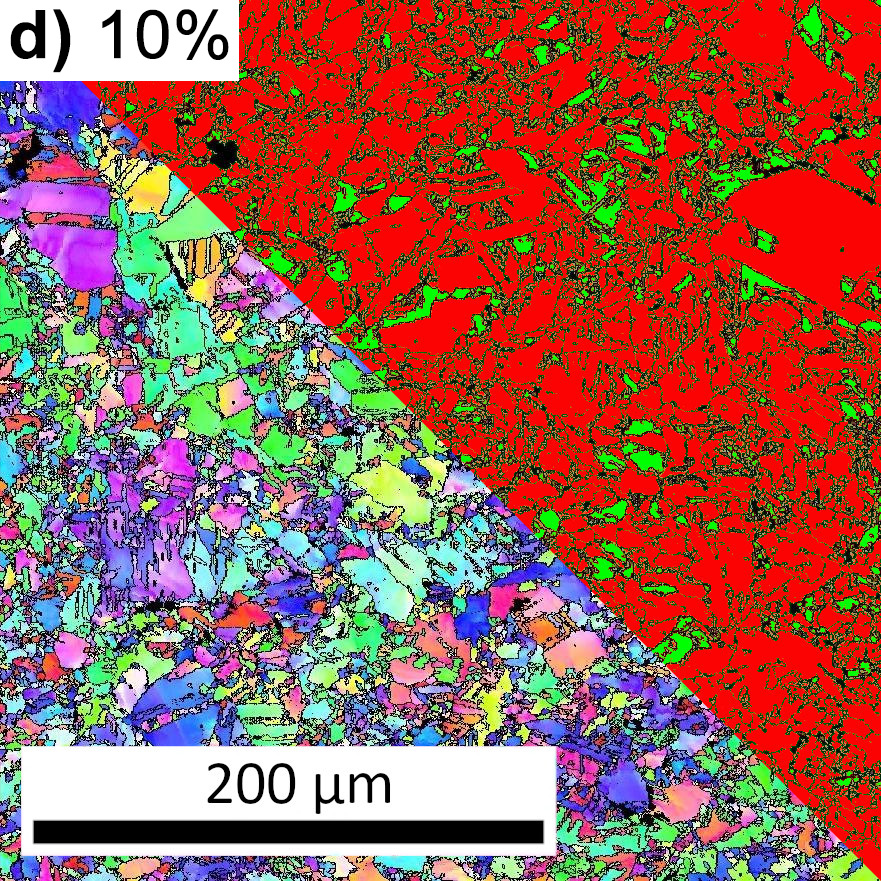}
    \end{subfigure}

    \vspace{0.2cm}

    \begin{subfigure}[b]{0.65\textwidth}
        \centering
        \includegraphics[width=\linewidth]{EBSD_legend_v3.png}
    \end{subfigure}

    \caption{Ex-situ EBSD observations (orientation maps and phase maps) of the co600 material during interrupted tests in compression: a) initial, b) \SI{1}{\percent} strain, c) \SI{4}{\percent} strain, d) \SI{10}{\percent} strain, and e) legend. Both the drawing and testing direction are parallel with the image horizontal axis.}
    \label{EBSD-ex600}
\end{figure}

\begin{figure}[]
\centering
    \begin{subfigure}[b]{0.25\textwidth}
        \centering
        \includegraphics[width=\linewidth]{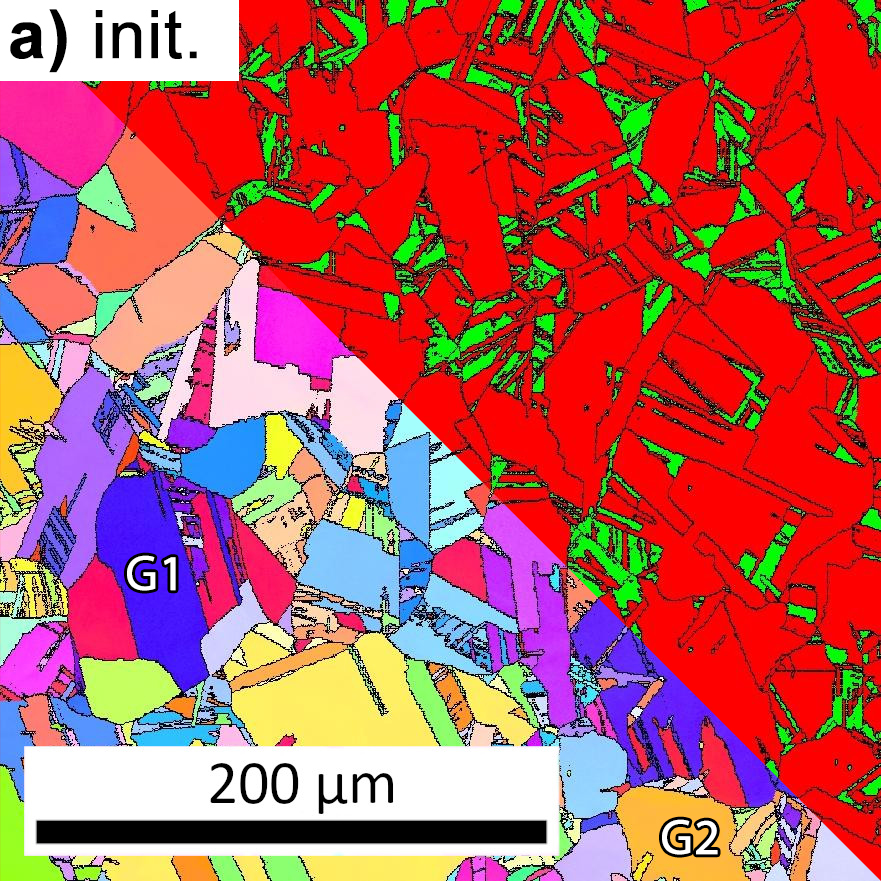}
    \end{subfigure} \hspace{1pt}
    \begin{subfigure}[b]{0.25\textwidth}
        \centering
        \includegraphics[width=\linewidth]{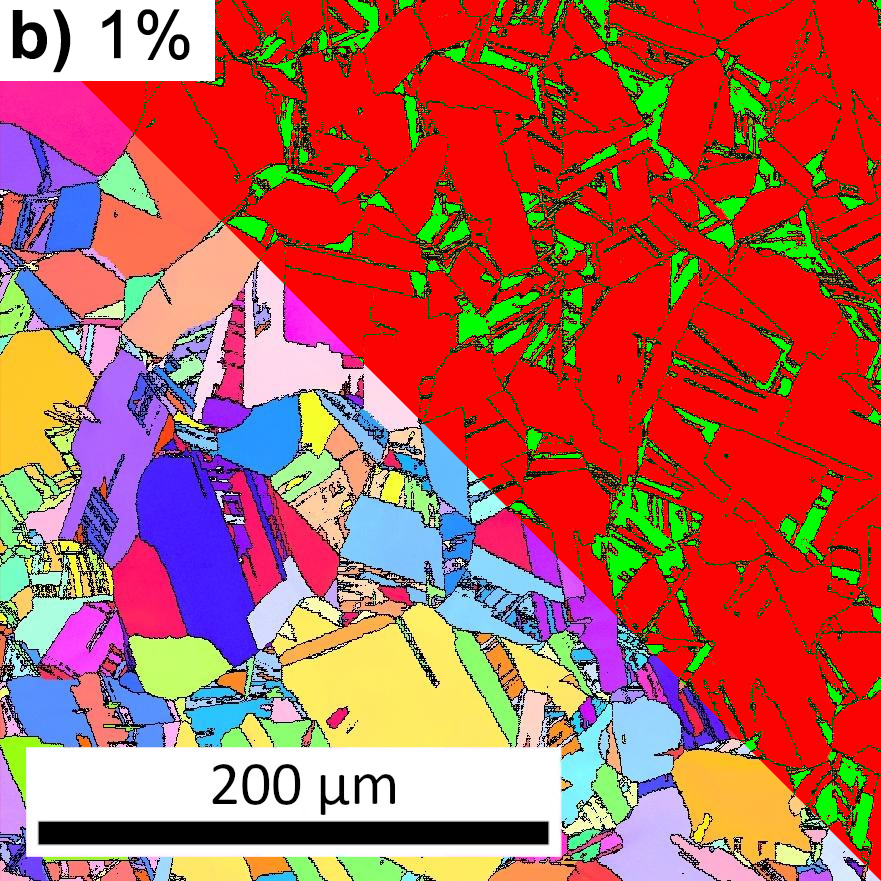}
    \end{subfigure}

    \vspace{0.2cm}

    \begin{subfigure}[b]{0.25\textwidth}
        \centering
        \includegraphics[width=\linewidth]{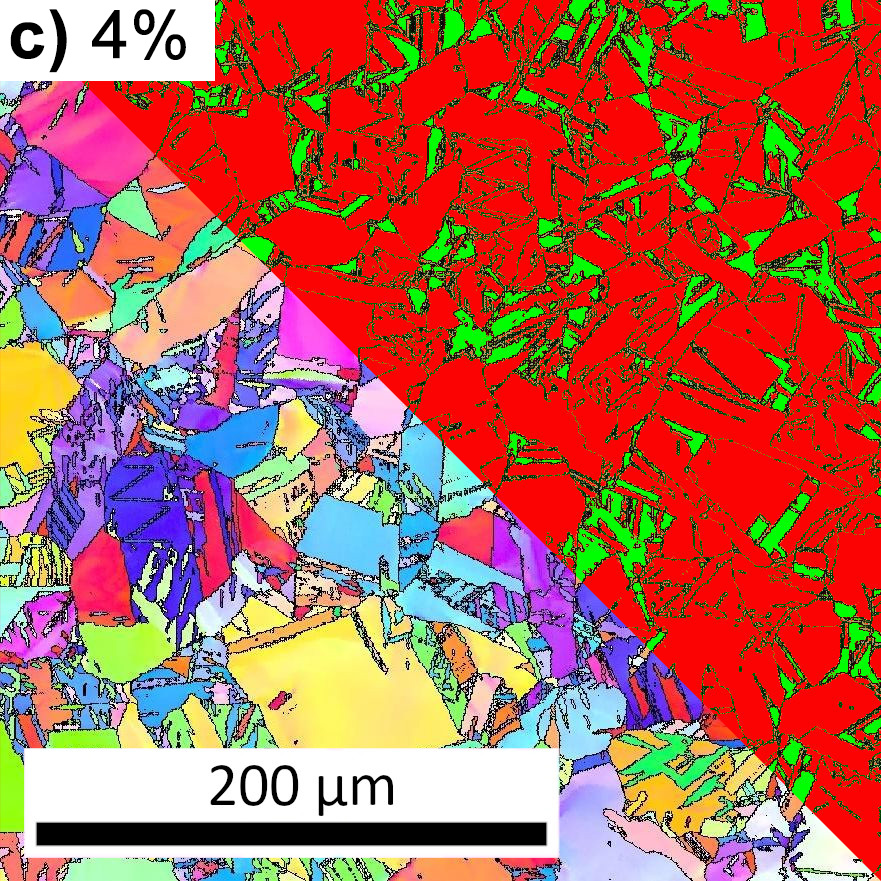}
    \end{subfigure} \hspace{1pt}
    \begin{subfigure}[b]{0.25\textwidth}
        \centering
        \includegraphics[width=\linewidth]{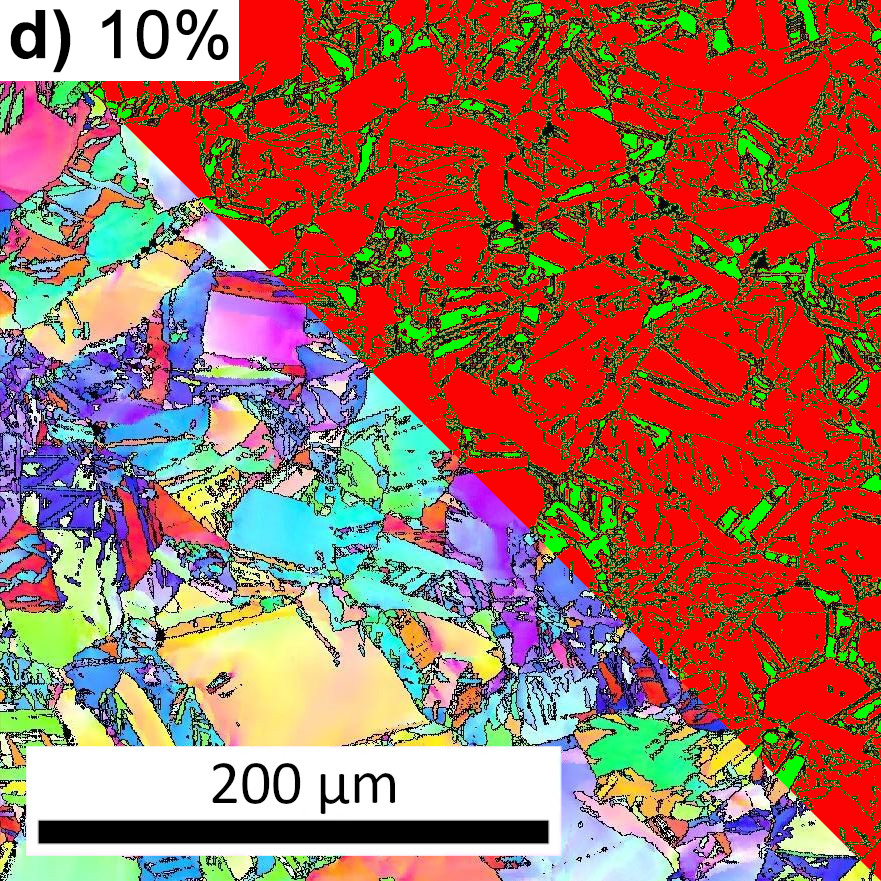}
    \end{subfigure}
    \caption{Ex-situ EBSD observations (orientation maps and phase maps) of the co800 material during interrupted tests in compression: a) initial, b) \SI{1}{\percent} strain, c) \SI{4}{\percent} strain, d) \SI{10}{\percent} strain.  Both the drawing and testing direction are parallel with the image horizontal axis. For legend see Fig. \ref{EBSD-ex600}.}
    \label{EBSD-ex800}
\end{figure}

\begin{figure}[]
\centering
    \begin{subfigure}[b]{0.25\textwidth}
        \centering
        \includegraphics[width=\linewidth]{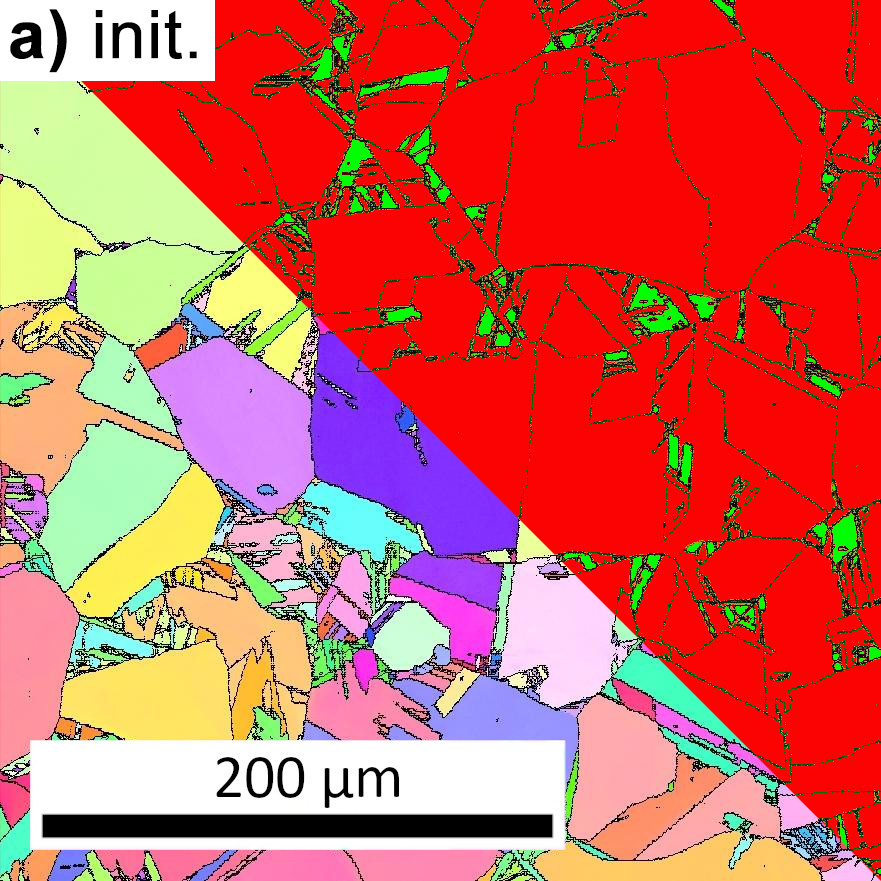}
    \end{subfigure} \hspace{1pt}
    \begin{subfigure}[b]{0.25\textwidth}
        \centering
        \includegraphics[width=\linewidth]{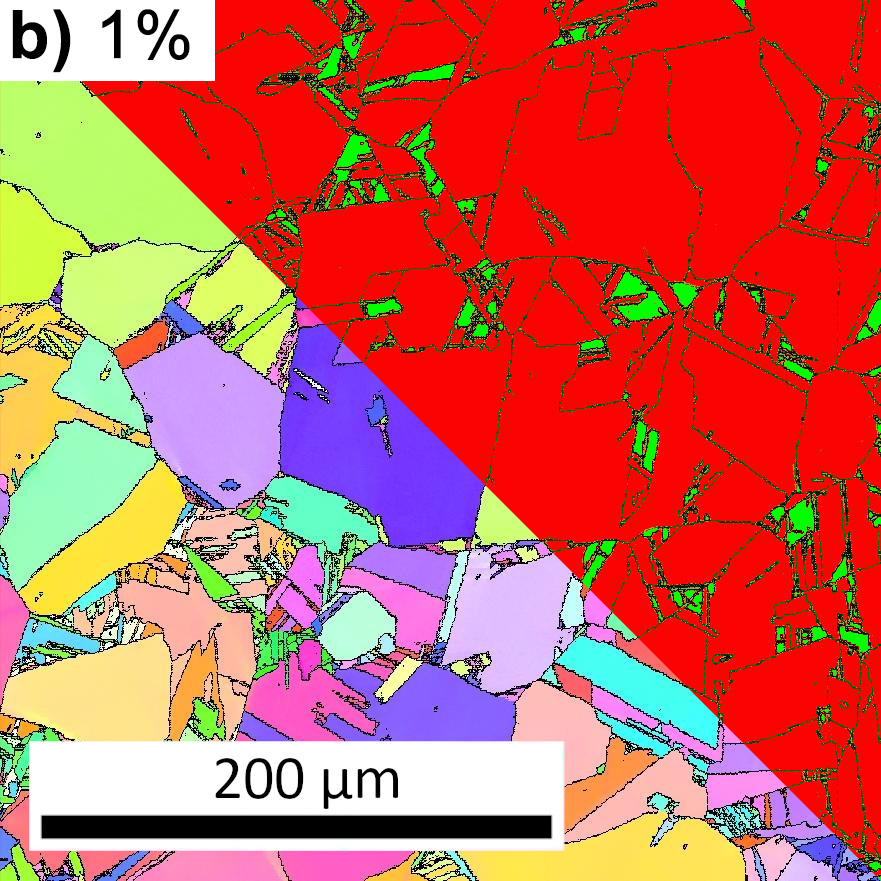}
    \end{subfigure}

    \vspace{0.2cm}

    \begin{subfigure}[b]{0.25\textwidth}
        \centering
        \includegraphics[width=\linewidth]{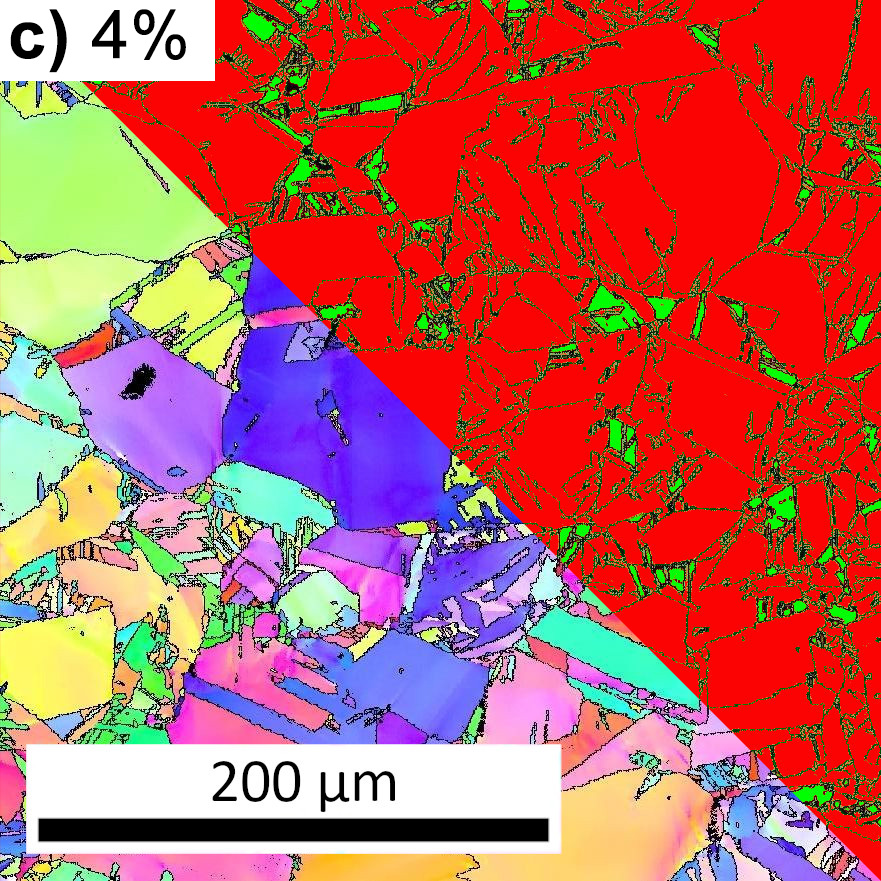}
    \end{subfigure} \hspace{1pt}
    \begin{subfigure}[b]{0.25\textwidth}
        \centering
        \includegraphics[width=\linewidth]{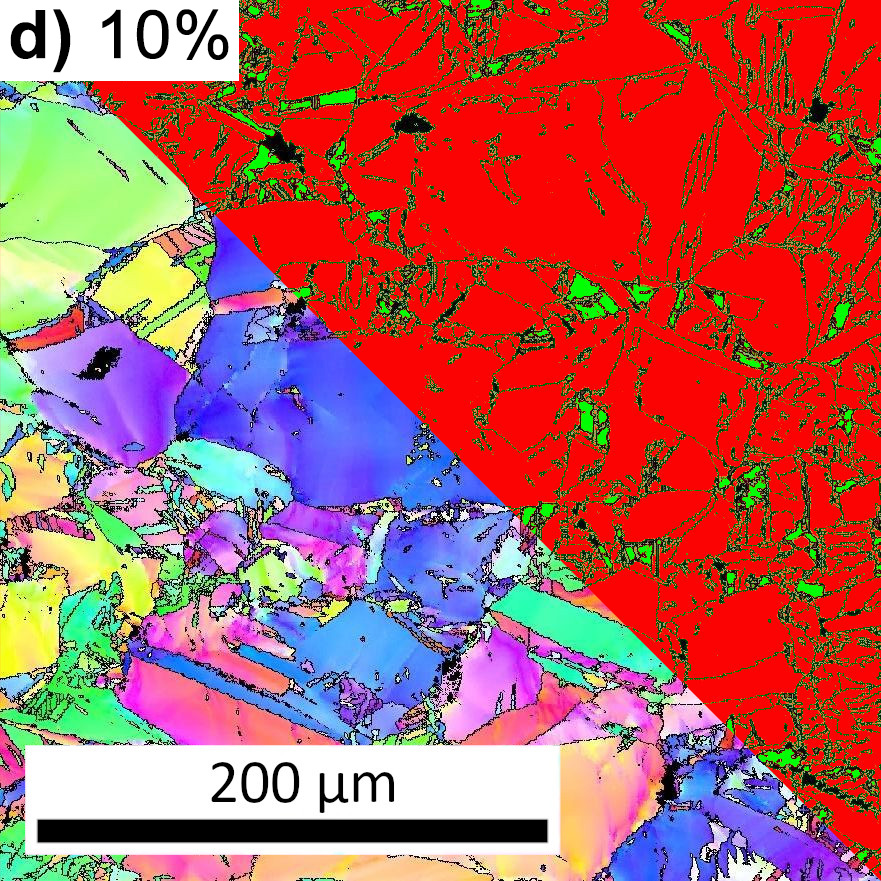}
    \end{subfigure}
    \caption{Ex-situ EBSD observations (orientation maps and phase maps) of the co1100 material during interrupted tests in compression: a) initial, b) \SI{1}{\percent} strain, c) \SI{4}{\percent} strain, d) \SI{10}{\percent} strain.  Both the drawing and testing direction are parallel with the image horizontal axis. For legend see Fig. \ref{EBSD-ex600}.}
    \label{EBSD-ex1100}
\end{figure}

\begin{figure}[]
\centering
    \begin{subfigure}[b]{0.25\textwidth}
        \centering
        \includegraphics[width=\linewidth]{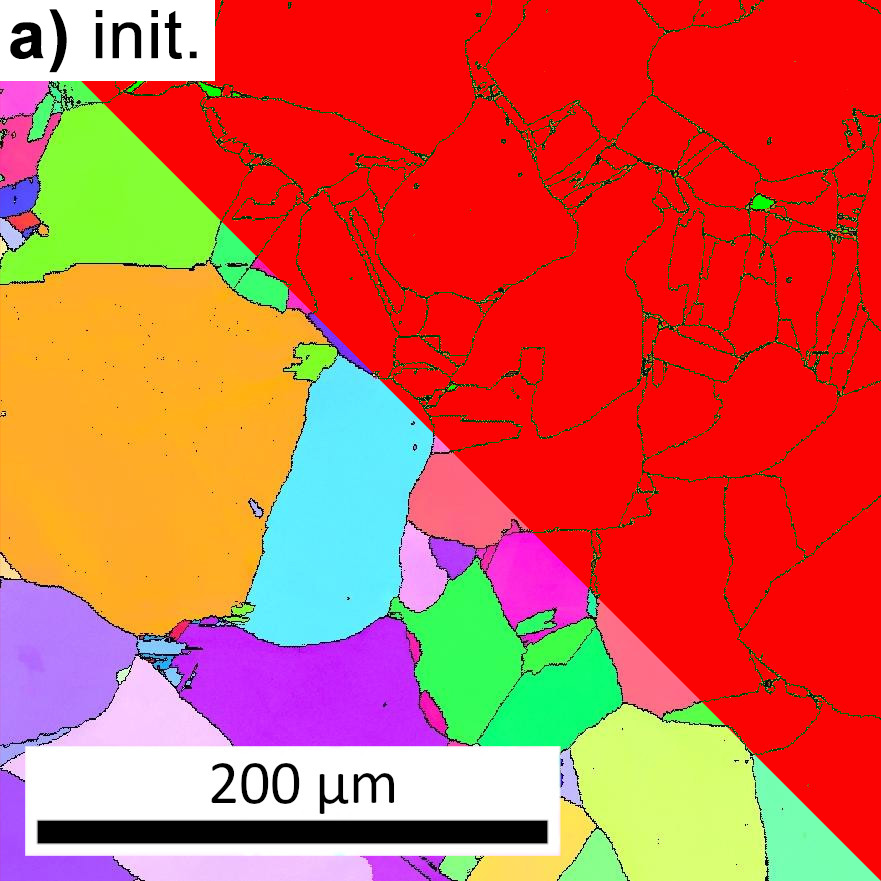}
    \end{subfigure} \hspace{1pt}
    \begin{subfigure}[b]{0.25\textwidth}
        \centering
        \includegraphics[width=\linewidth]{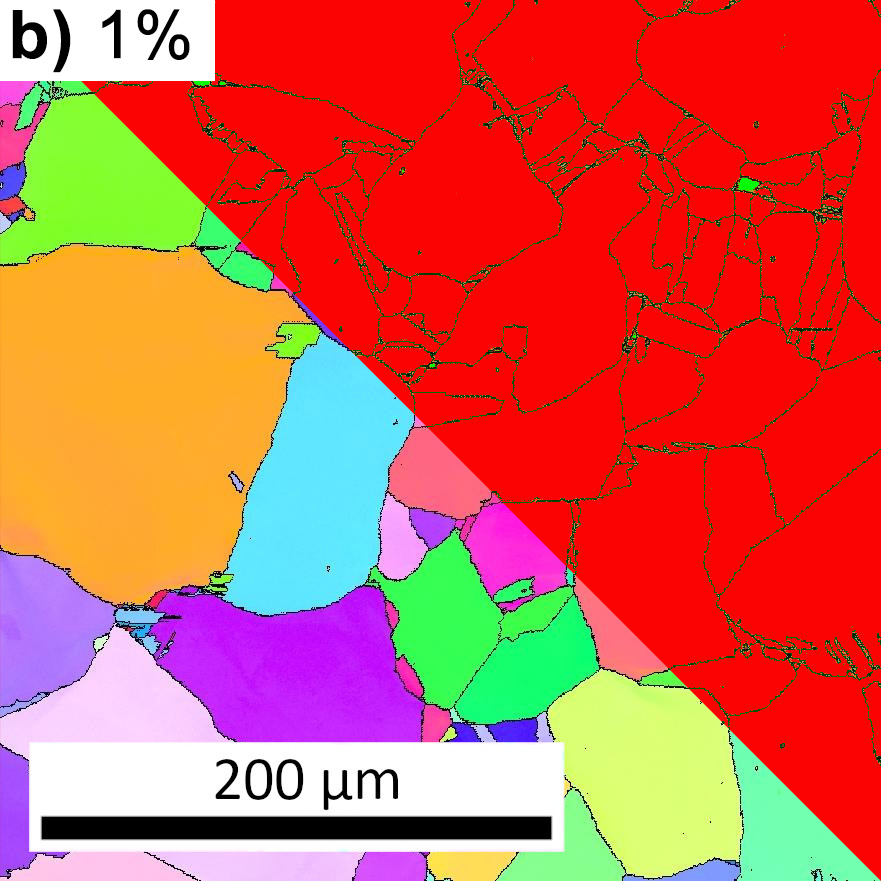}
    \end{subfigure}

    \vspace{0.2cm}

    \begin{subfigure}[b]{0.25\textwidth}
        \centering
        \includegraphics[width=\linewidth]{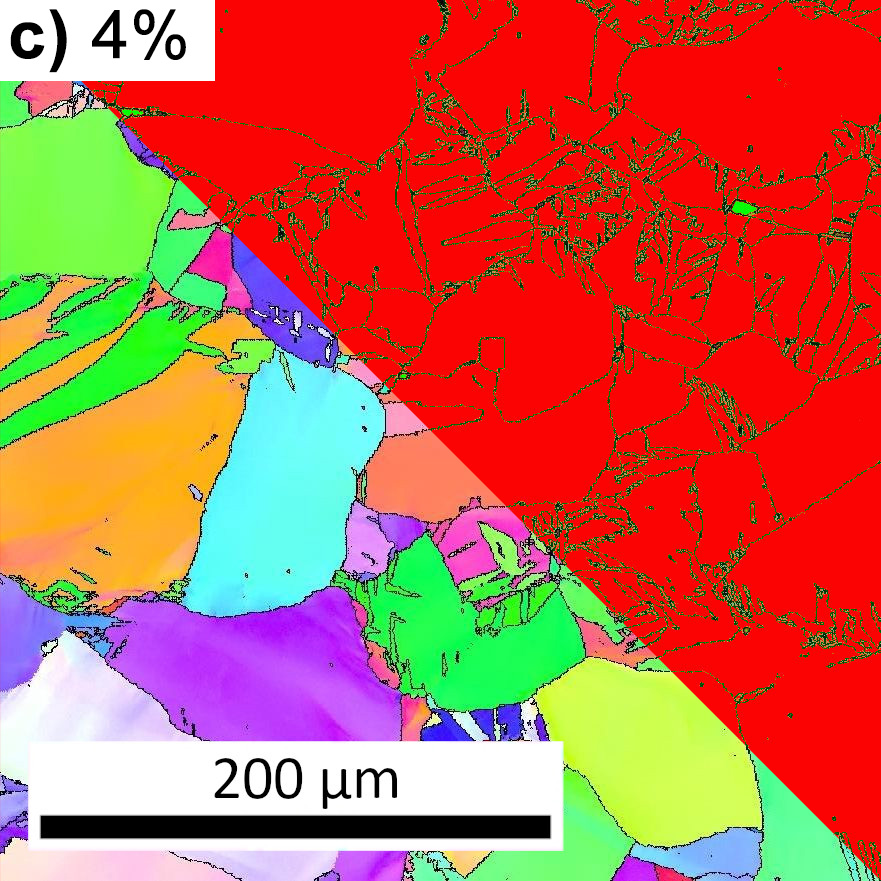}
    \end{subfigure} \hspace{1pt}
    \begin{subfigure}[b]{0.25\textwidth}
        \centering
        \includegraphics[width=\linewidth]{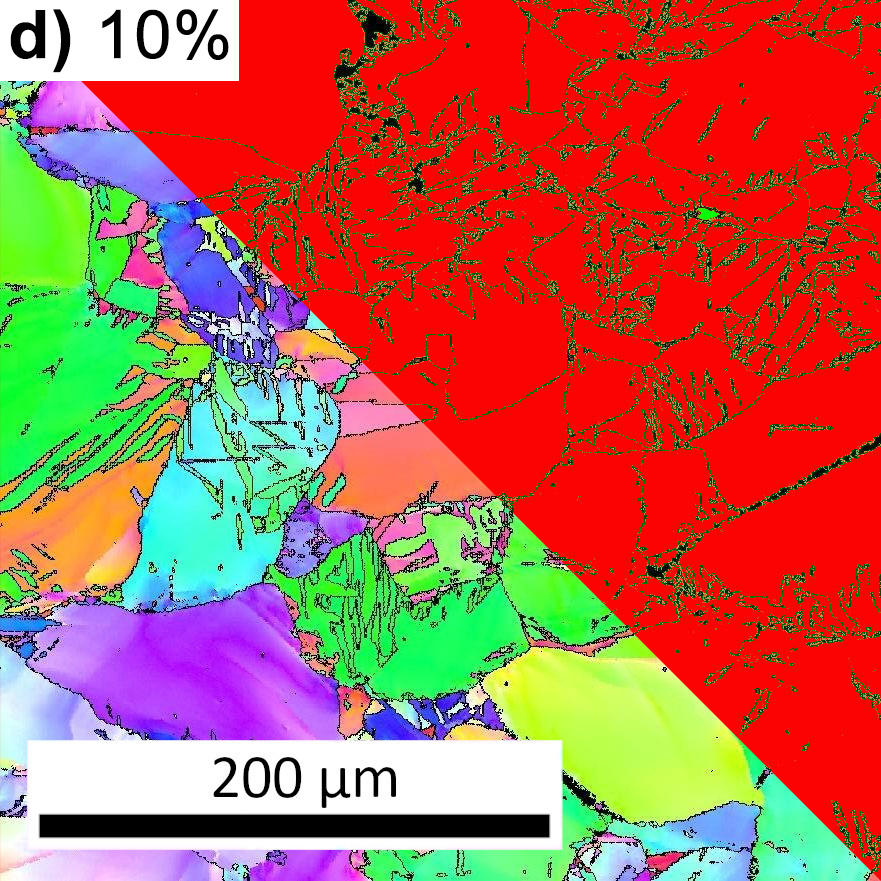}
    \end{subfigure}
    \caption{Ex-situ EBSD observations (orientation maps and phase maps) of the co1100-20c material during interrupted tests in compression: a) initial, b) \SI{1}{\percent} strain, c) \SI{4}{\percent} strain, d) \SI{10}{\percent} strain.  Both the drawing and testing direction are parallel with the image horizontal axis. For legend see Fig. \ref{EBSD-ex600}.}
    \label{EBSD-ex1100-20c}
\end{figure}

\begin{figure}[]
\centering
    \begin{subfigure}[b]{0.45\textwidth}
        \centering
        \includegraphics[width=\linewidth]{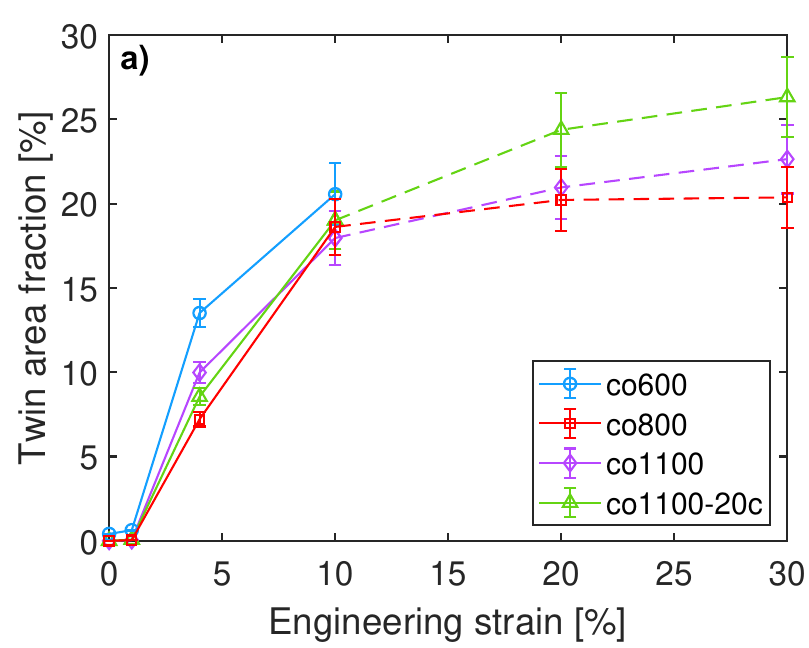}
    \end{subfigure} \hspace{1pt}
        \begin{subfigure}[b]{0.45\textwidth}
        \centering
        \includegraphics[width=\linewidth]{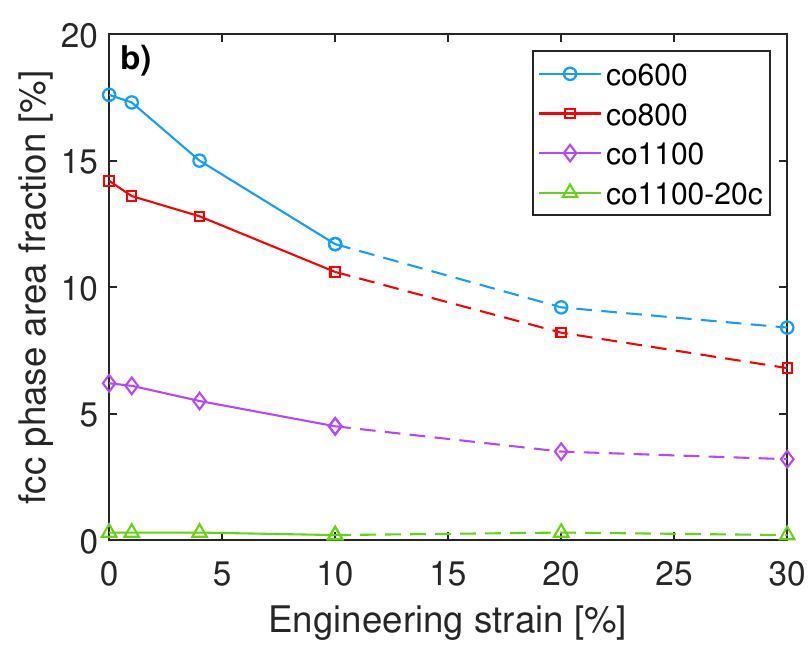}
    \end{subfigure}
    \caption{Quantitative data of the area fraction of a) the mechanical twins and b) the fcc phase calculated from the ex-situ EBSD observations of all the investigated materials.}
    \label{EBSD-quant}
\end{figure}

On that account, the materials were also investigated by means of deformation tests interrupted at different strain levels in order to acquire EBSD data (Fig. \ref{EBSD-ex600}--\ref{EBSD-ex1100-20c}). EBSD was performed in the initial pristine condition and then in the same area at \SI{1}{\percent}, \SI{4}{\percent}, and \SI{10}{\percent} engineering strain (cf. Fig. \ref{def}). The acquired EBSD data were also used for quantitatively examining the evolution of (i) the twin area fraction and (ii) the fcc phase area fraction with increasing strain. Fig. \ref{EBSD-quant} shows the results of these analyses with the aim of shedding more light on the effects observed in the compression curves and AE results. It must be noted that while EBSD was measured also at \SI{20}{\percent} and \SI{30}{\percent} strain, these results were somewhat less reliable due to a high number of low CI points in the EBSD maps and thus must be taken with caution and attention must paid to the trends rather than to the absolute values (this holds especially for the co600 material, where its fine not fully recrystalized microstructure combined with accumulating deformation rendered the EBSD maps at \SI{20}{\percent} and \SI{30}{\percent} strain less accurate at given conditions; for these ancillary EBSD micrographs see \cite{gres_experimental_2023}). Hence, the trends beyond \SI{10}{\percent} strain in Fig. \ref{EBSD-quant} are marked with dashed lines. 

The orientation maps of all the materials (Fig. \ref{EBSD-ex600}--\ref{EBSD-ex1100-20c}) share some common features and bear witness of accumulating deformation in terms of dislocation slip activity (i.e. developing lattice curvature manifested by means of color gradients within grains) as well as the activation of $\{ 10 \Bar{1} 2 \}$-type twinning. Initially, the twins in each material are mostly thin and, subsequently, many of the twins grow laterally upon further deformation, sometimes completely consuming the parent grain (see the example grains marked G1 and G2 in Fig. \ref{EBSD-ex800}). Figure \ref{EBSD-quant}a shows the evolution of twin area fraction during deformation. The initial condition and the condition at \SI{1}{\percent} strain (i.e. slightly before the yield point, cf. Fig \ref{def}) show very few or no twins. Irrespective of the material, most twin nucleation occurs between the \SI{1}{\percent} and \SI{10}{\percent} strain, with few or no new twins appearing after \SI{10}{\percent}. The increase in the twin area fraction between \SI{4}{\percent} and \SI{10}{\percent} strain is a combination of twin nucleation and growth and its further increase beyond \SI{10}{\percent} strain can be mostly attributed to twin growth. It is worth noting that twin nucleation, appearing mostly around the yield point, happens to correlate with the recorded AE activity (cf. Fig \ref{AE}). This observation will be further elaborated in the Discussion.

The evolution of the fcc fraction presented in Fig. \ref{EBSD-quant}b shows a gradual decrease during deformation in each sample. The rate at which the fcc phase fraction decreases is higher in the case of higher initial values. Note that the higher spread in the initial fcc phase content across the samples results from a more local nature of EBSD in this analysis compared to much larger maps used for the primary material characterization, as reported above and in \cite{knapek_effect_2020,gres_experimental_2023}. Nonetheless, it is evident that during loading the transformation slows down and the samples still contain a considerable fraction of the original fcc phase at \SI{30}{\percent} strain. The stress-induced fcc$\rightarrow$hcp transformation thus appears gradual and rather sluggish, and does not seem to be reflected in the AE activity (which is absent except for the vicinity of the yield point), as shown in Fig. \ref{AE}.

\section{Discussion}

\begin{figure}[b]
\centering
    \begin{subfigure}[b]{0.24\textwidth}
    \centering
        \includegraphics[width=\linewidth]{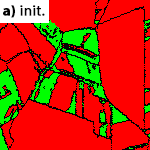}
    \end{subfigure}
    \begin{subfigure}[b]{0.24\textwidth}
        \centering
        \includegraphics[width=\linewidth]{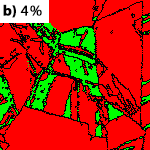}
    \end{subfigure}
    \begin{subfigure}[b]{0.24\textwidth}
        \centering
        \includegraphics[width=\linewidth]{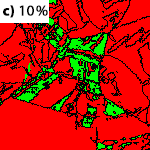}
    \end{subfigure}
    \begin{subfigure}[b]{0.24\textwidth}
        \centering
        \includegraphics[width=\linewidth]{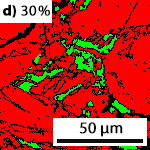}
    \end{subfigure}
    \caption{Close-up of the ex-situ EBSD phase maps (green -- fcc, red -- hcp) of the co800 material a) in the initial condition, and at the strain of b) \SI{4}{\percent}, c) \SI{10}{\percent}, and d) \SI{30}{\percent}, showing the evolution of shape and size of the fcc grains.}
    \label{MT-dynamics}
\end{figure}

The compression tests (Fig. \ref{def}) showed some dependence of the yield point, $\sigma_{0.2}$, on the annealing temperature for the samples co600, co800, and co1100 not subjected to thermal cycling. This behavior reflects the well-known connection with the grain size and was observed in cobalt before \cite{fleurier_size_2015, hug_size_2019}, even though the fcc phase fraction can, to some extent, modulate this effect. On the other hand, the differences are very slight and it can be inferred that the $\sigma_{0.2}$ value rather saturates quite rapidly beyond annealing at \SI{800}{\celsius}. Similarly, the thermally cycled co1100-20c showed only a slight decrease in $\sigma_{0.2}$, in line with the stabilization of the grain size (Fig. \ref{GS}) and possible effect of the disappearance of fcc/hcp grain boundaries (Fig. \ref{EBSD_TT}). Also, the value of fracture strain, $\varepsilon_{\mathrm{max}}$, seems to increase from sample co600 to samples co800 and co1100. Likely explanation is again the recrystallization accelerated by higher annealing temperature. The recrystallization is also reflected in the disappearance of crystallographic texture after annealing (cf. Fig. \ref{EBSD_init} and \ref{EBSD_TT}). Interestingly, after the thermal cycling (sample co1100-20c) the value of $\varepsilon_{\mathrm{max}}$ drops significantly (Fig. \ref{def}), although the grain size still somewhat increases (Fig. \ref{GS}). This can be related to the reduced residual fcc phase (\SI{<0.5}{\percent}) in this material compared to several per cent in co600, co800, and co1100, as was best shown in Fig \ref{EBSD_TT}. While the fcc$\rightarrow$hcp transformation brings about only \tld{}\SI{0.4}{\percent} volume change, the abundance of slip systems in the fcc structure can accommodate considerable shape changes within the fcc grains \cite{sanderson_deformation_1972, dubos_temperature_2020}. The transformation and slip are interrelated and, together, were shown to be capable of accommodating several per cent of plastic strain \cite{sanderson_deformation_1972}. Such an explanation is also supported by the interrupted deformation tests, as revealed in Fig. \ref{MT-dynamics} by means of zoomed-in ex-situ EBSD phase maps of the co800 material up to \SI{30}{\percent} strain: the fcc grains significantly change shape in the course of deformation whereas the transformation to hcp proceeds rather gradually. Also, the presence of the fcc phase can contribute to the ultimate strength, $\sigma_{\mathrm{max}}$, (cf. Fig. \ref{def}) as the intersecting slip planes bring about additional strain hardening \cite{dieter_mechanical_1988}.

The twin fraction evolution during compression for all the materials adhered to a pattern of pronounced increase caused by abundant nucleation up to \SI{10}{\percent} strain, followed by slowing down of the rate of the twin area fraction increase at higher strain when twin nucleation was exhausted and, primarily, twin growth took place (Fig. \ref{EBSD-ex600}--\ref{EBSD-ex1100-20c}). Some twins are observed to nucleate as soon as at \SI{1}{\percent} strain, i.e. within the apparently elastic stage of deformation. This observation together with the AE results (Fig. \ref{AE}) suggest that some irreversible plastic changes take place already before macroscopic yielding (note that some contribution from collective dislocation activity can also be expected). Such a quasi-elastic stage of deformation exhibiting signs of (micro)plasticity was observed in hcp metals before \cite{chmelik_exploring_2012,maas_micro-plasticity_2018}. The interrupted deformation experiments testified that the evolution of twinning dynamics during deformation agrees well with the observed AE signals: high AE amplitudes appear when new twins are being nucleated and AE is significantly reduced in the later stages of compression accompanied by twin growth. Lateral growth of existing twins is a much less rapid process than twin nucleation and is known to be a very weak source of AE \cite{chmelik_exploring_2012}. Increasing amplitudes of the AE signals around elasto-plastic transition with more extensive thermal treatment (Fig. \ref{AE}) most likely appear due to accelerated recrystallization: bigger resulting grains provide more room for twin tip propagation during nucleation as well as longer mean free path for dislocation avalanche motion, especially when also the residual fcc phase has vanished. On the other hand, the fraction of the fcc phase continues to decrease in all the samples beyond the region where AE is recorded. This observation manifests that the transformation is rather sluggish (at least at the experimental conditions used in this study) even though it was documented that the free energy change upon this stress-induced transition is roughly an order of magnitude higher than in the case of thermal cycling \cite{sanderson_deformation_1972}. The EBSD results support this notion as the fcc grains were shown to  decrease in size during loading gradually (Fig. \ref{MT-dynamics}). Such a continuous transformation lacks rapid dynamics needed to generate detectable AE signals and is in contrast with burst-like martensitic transformation increments that give rise to pronounced AE observed in many metallic materials \cite{voronenko_acoustic_1982, manosa_acoustic_1990, toth_acoustic_2020}.

\begin{figure}[]
\centering
    \begin{subfigure}[b]{0.32\textwidth}
        \centering
        \raisebox{0.038\height}{\includegraphics[width=\linewidth]{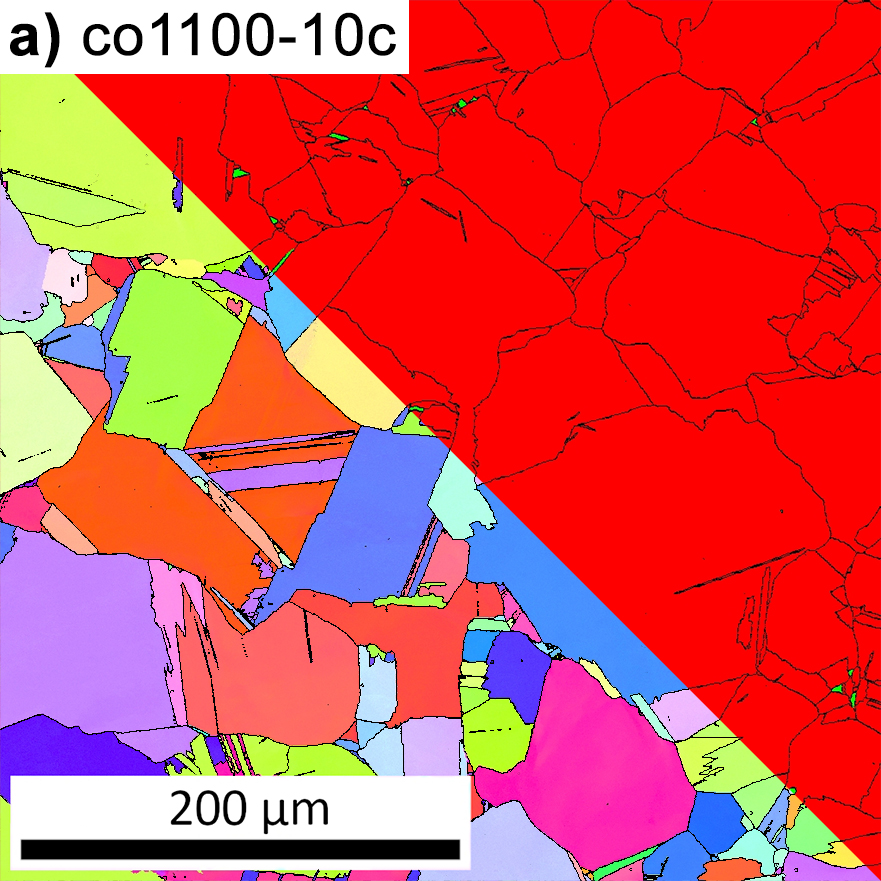}}
    \end{subfigure} \hspace{8pt}
    \begin{subfigure}[b]{0.43\textwidth}
        \centering
        \includegraphics[width=\linewidth]{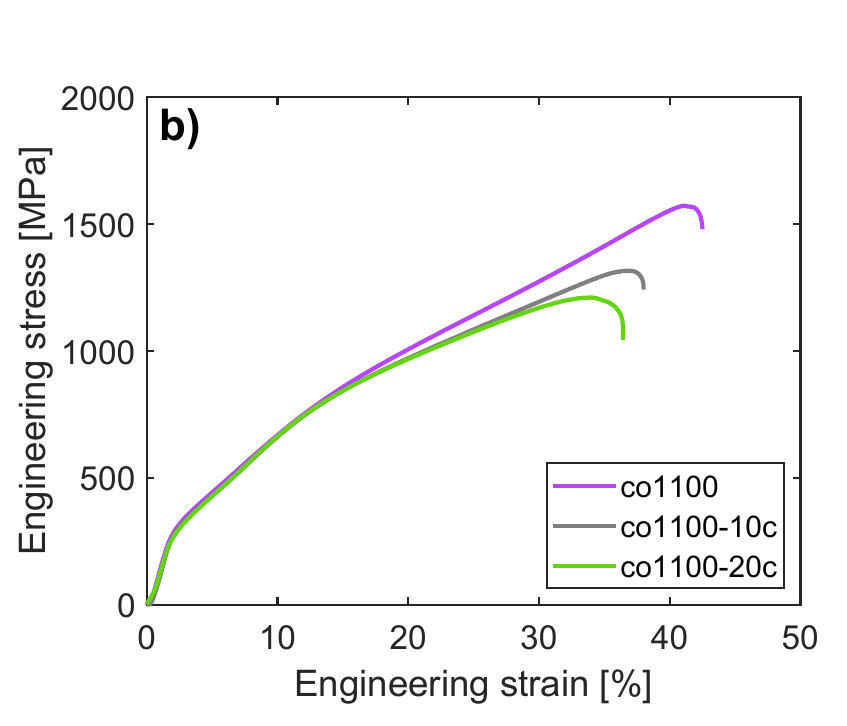}
    \end{subfigure}
    \caption{a) Microstructure of the co1100-10c material used for supplementary analyses. For legend see Fig. \ref{EBSD-ex600}. b) Comparison of the compression curves of the co1100, co1100-10c, and co1100-20c materials.}
    \label{def_1100}
\end{figure}

In order to gather further insights into the effect of thermal cycling on the deformation dynamics, the co1100-10c material (i.e. co1100 subjected to only 10 thermal cycles) was further examined as an intermediate between co1100 and co1100-20c. The microstructure of co1100-10c is shown in Fig. \ref{def_1100}a and, at first glance, exhibits no obvious differences compared to co1100-20c (cf. Fig. \ref{EBSD_TT}). The co1100-10c material was tested in compression and the curve was compared to those of co1100 and co1100-20c, as presented in Fig \ref{def_1100}b. In terms of $\varepsilon_{\mathrm{max}}$ and $\sigma_{\mathrm{max}}$ the curve falls impeccably between the co1100 and co1100-20c ones despite the fact that the fcc phase was mostly eliminated (\SI{<1}{\percent}) in both co1100-10c and co1100-20c samples and the determined grain size was also practically identical (\tld$\SI{65}{\um}$); cf. Fig. \ref{GS}, \ref{EBSD_TT}, \ref{def_1100}. It is important to note that similar trends were observed also after thermal cycling of samples annealed at lower temperatures (\SIrange{600}{1000}{\celsius}) \cite{knapek_effect_2020,gres_experimental_2023}. To explain this phenomenon, let us first endeavor to consider a traditional explanation in terms of the grain structure and the activity of deformation mechanisms. It was shown that larger grain size in hcp materials can cause twinning to be activated more easily \cite{dobron_grain_2011, bohlen_effect_2006}. The high angle grain boundaries (HAGBs) introduced by twinning present new barriers to dislocation slip that can lead to the deteriorated deformability in materials with larger grains \cite{wang_deformation_2014, krajnak_influence_2019}. At the same time, higher twinned volume can enhance (basal) slip activity because of changes in the crystal orientation, opposing the effect of decreasing $\varepsilon_{\mathrm{max}}$ \cite{drozdenko_investigating_2016}. However, there was only a minor grain size increment between co1100 and co1100-20c (\tld$\SI{50}{\um}$ vs. \tld$\SI{65}{\um}$) and the interrupted deformation tests showed only a slight increase in the twinned fraction in co1100-20c compared to co1100 and, in fact, to all the annealed samples (Fig. \ref{EBSD-quant}a). Moreover and most importantly, the explanation related to grain size cannot account for the differences detected in the deformation curves of co1100-10 and co1100-20 (both having rather identical grain size and negligible fcc fraction). Thus, the resolution must lie in different microstructural characteristics and, indeed, is most likely related to the \emph{type} and not the absolute amount of grain boundaries. 

\begin{figure}[t]
\centering
    \begin{subfigure}[b]{0.25\textwidth}
        \centering
        \includegraphics[width=\linewidth]{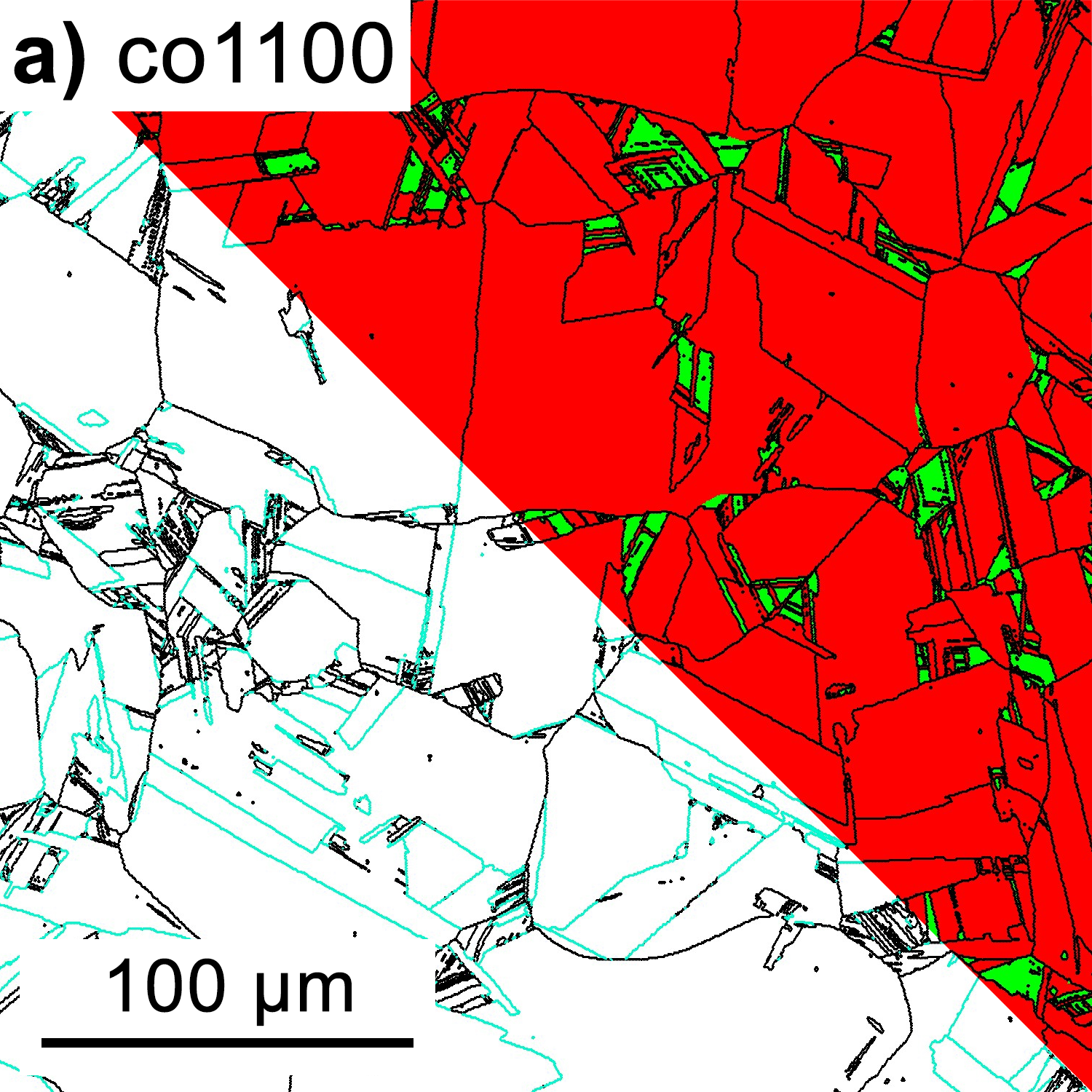}
    \end{subfigure} \hspace{1pt}
    \begin{subfigure}[b]{0.25\textwidth}
        \centering
        \includegraphics[width=\linewidth]{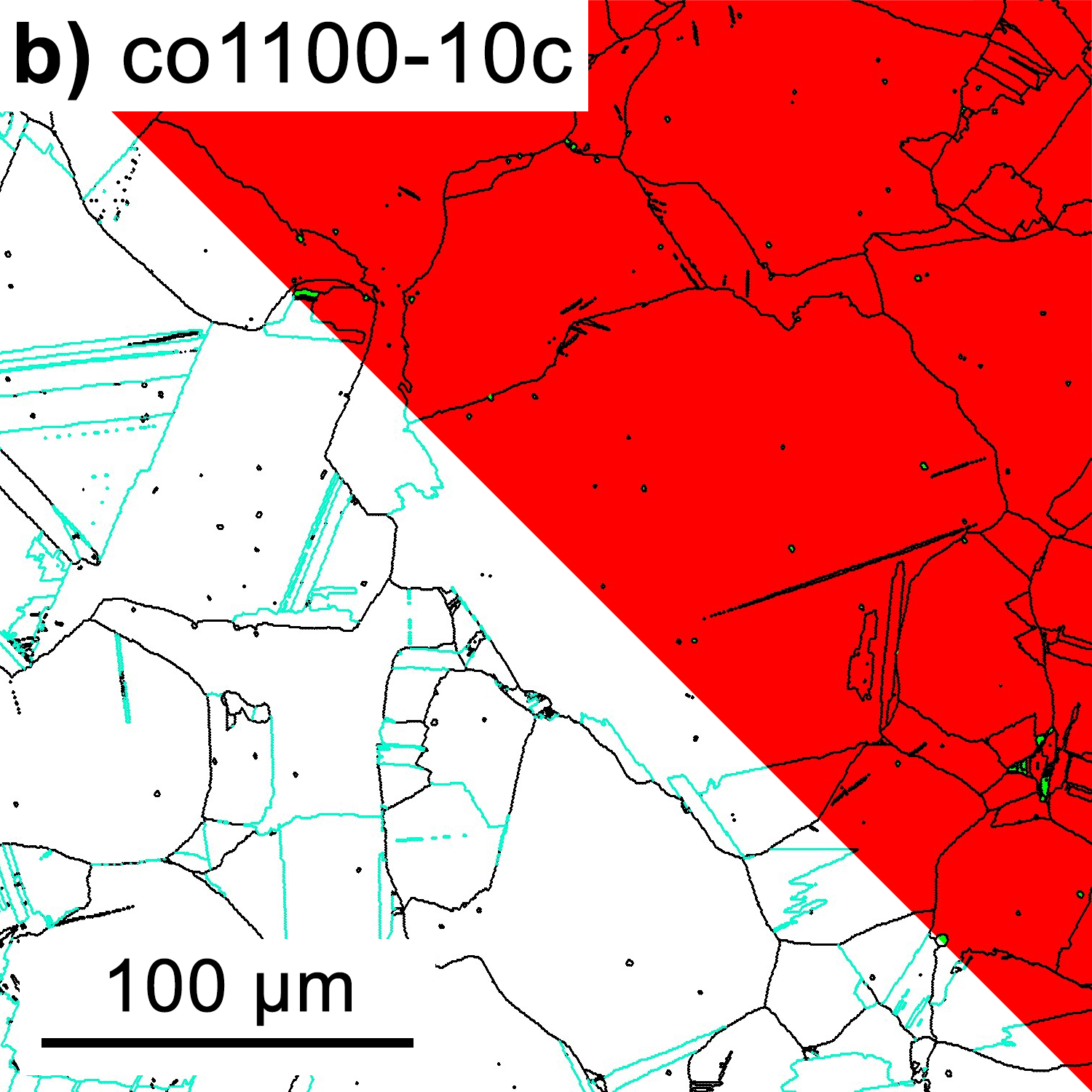}
    \end{subfigure} \hspace{1pt}
    \begin{subfigure}[b]{0.25\textwidth}
        \centering
        \includegraphics[width=\linewidth]{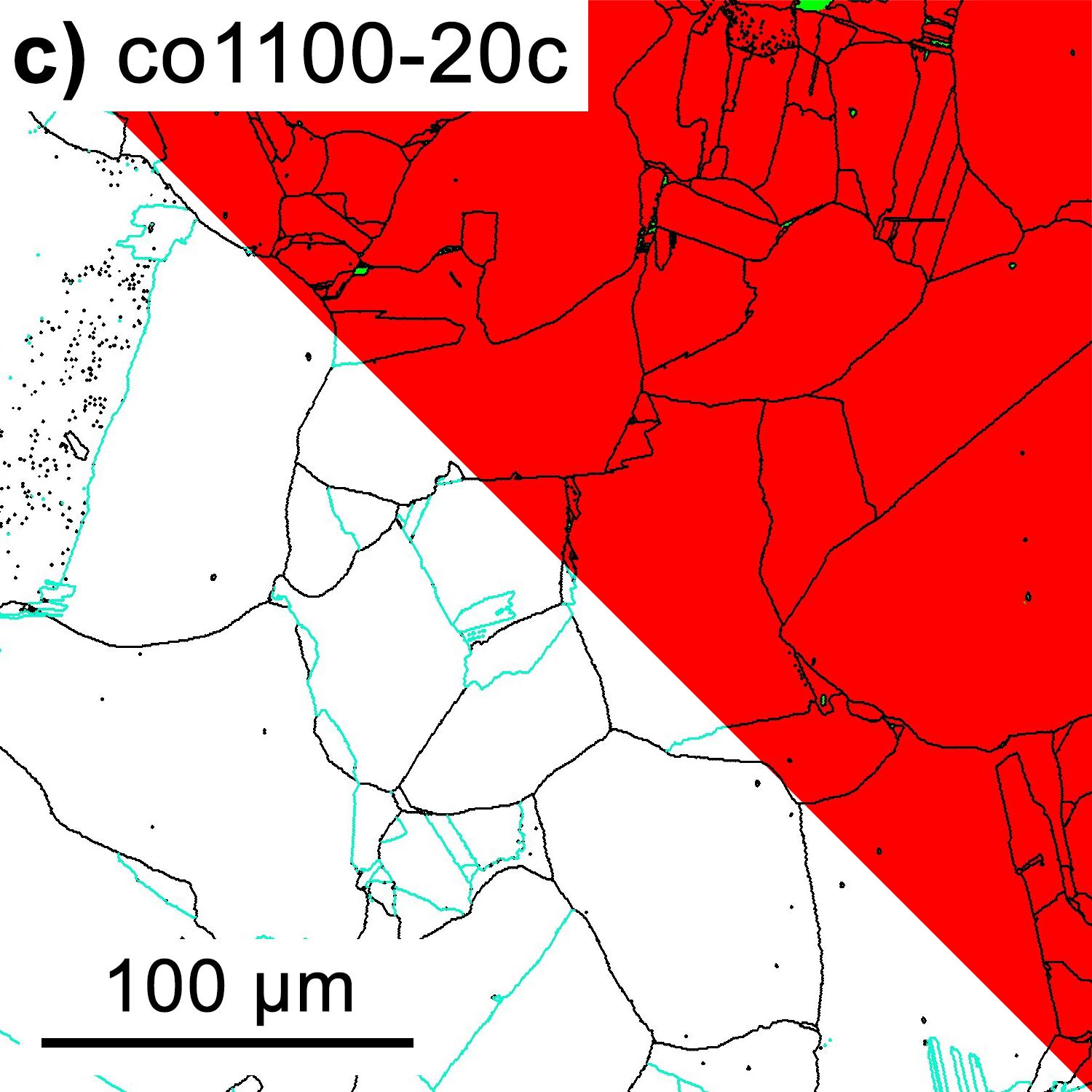}
    \end{subfigure}
    
    \vspace{0.2cm}
    
    \begin{subfigure}[b]{0.82\textwidth}
        \centering
        \includegraphics[width=\linewidth]{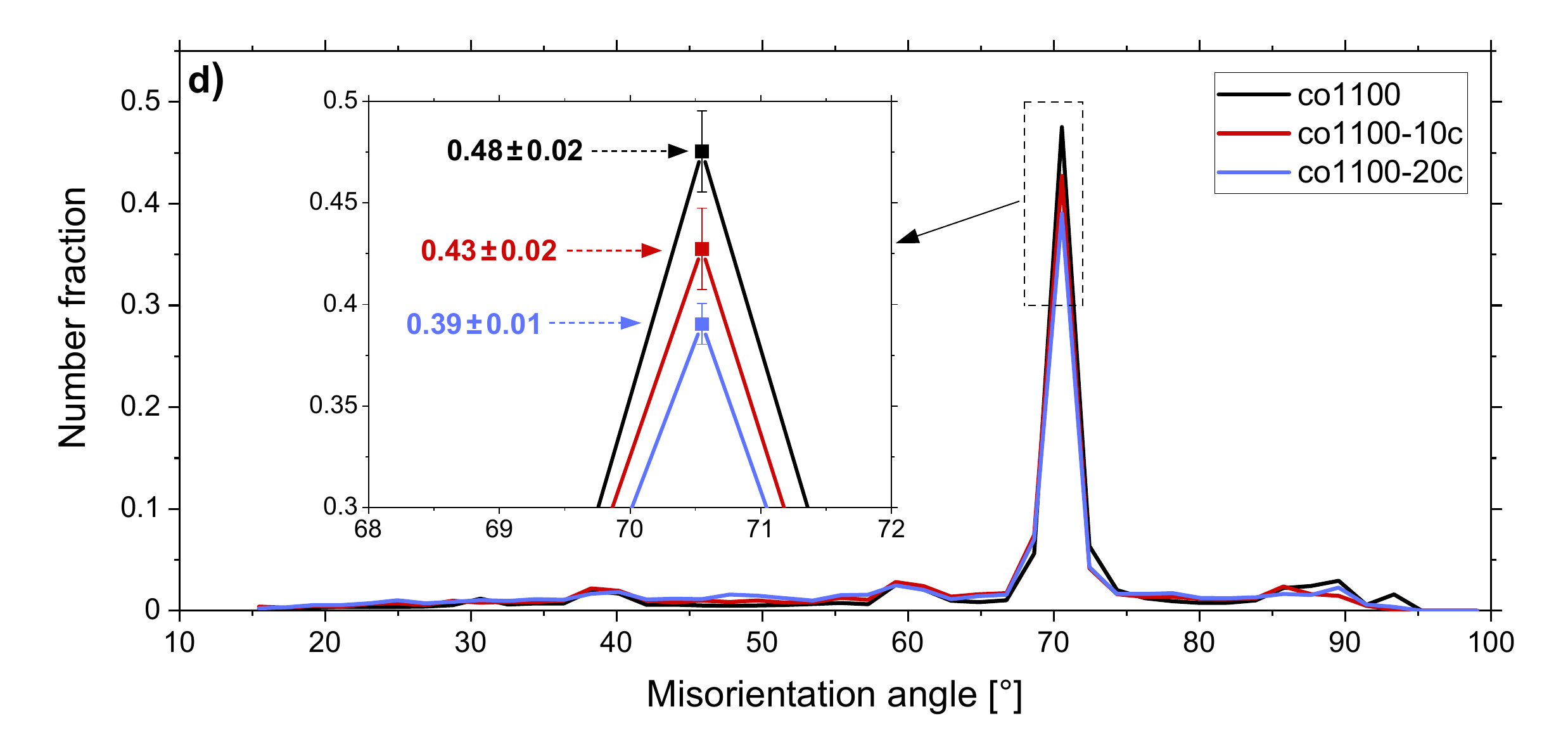}
    \end{subfigure}
    \caption{Detail of the grain structures observed in the a) co1100, b) co1100-10c, and c) co1100-20c materials by means of diagonally split EBSD maps showing grain boundary types (left bottom; black -- regular HAGBs, cyan -- special \tld{}\ang{71} HAGBs) and phase maps (green -- fcc, red -- hcp). d) HAGBs misorientation angle distribution for the co1100, co1100-10c, and co1100-20c materials calculated from large EBSD maps (two 1000$\times$\SI{1000}{\um^2} scans of different areas, step size \SI{2}{\um}; error bars show the scans variance). The \tld{}\ang{55} fcc/hcp HAGBs are excluded from the analyses to enable direct comparison of the \tld{}\ang{71} HAGBs fractions.}
    \label{GB}
\end{figure}

While the first ten cycles were shown to eliminate the residual fcc phase almost completely, the reported thermal analyses suggest a further evolution of the transformation (and the resulting microstructure) between 10 and 20 cycles \cite{bauer_kinetics_2011,knapek_effect_2020}. This is connected to the reorganization of dislocation structures taking part in the transformation and the related evolution of shape and size of the hcp variants (``martensite plates'') within grains during transformation, es elaborated in detail in \cite{bauer_kinetics_2011}. As mentioned above, there are, indeed, four hcp martensite variants corresponding to four equivalent \{111\} planes in the fcc phase. Geometrically, taking into account the Shoji-Nishiyama relation, the misorientation angle between such variants is \tld{}\ang{71} (whereas the misorientation between hcp and fcc is \tld{}\ang{55}) \cite{nishiyama_crystallography_1978, toledano_theory_2001}. Fig.~\ref{GB} shows the analysis of such boundaries in the undeformed samples co1100, co1100-10c, and co1100-20c (i) visually in Fig. \ref{GB}a--c and (ii) quantitatively in terms of the fraction of such boundaries in Fig \ref{GB}d. Note that due to the fact that the \tld{}\ang{55} boundaries are fathomably present only in the co1100 sample, they are excluded from the analysis to allow for a direct comparison of the fraction of \tld{}\ang{71} boundaries with the other two materials. It is observed that the \tld{}\ang{71} boundaries amount to roughly half of all HAGBs in co1100 and with the increasing number of thermal cycles (i) their quantity drops and (ii) they appear less rectangular. This grain boundary evolution is in agreement with the model proposed in \cite{bauer_kinetics_2011} asserting that the dislocation structures inside the parent hcp grains in cobalt evolve (while gradually pushing the unfavorable dislocation stacks into the non-\ang{71} HAGBs) and, after many cycles, eventually stabilize such that they facilitate a complete transformation of the fcc grain into one hcp martensite variant. The EBSD data were further evaluated to assess the Kernel Average Misorientation (KAM) that quantifies local lattice curvatures and scales with dislocation density in the material. The mean KAM angle increased with thermal cycling as follows: co1100 –- \tld{}\ang{0.33}, co1100-10c -- \tld{}\ang{0.39}, co1100-20c –- \tld{}\ang{0.43}, thus giving evidence to rising cumulative length of dislocation arrays (involved in the realization of hcp$\leftrightarrow$fcc transformation) upon thermal cycling, again in agreement with \cite{bauer_kinetics_2011}.  

The \tld{}\ang{71} boundaries are considered \emph{special} within the concept of grain boundary engineering proposed by Watanabe \cite{watanabe_grain_2011}. Owing to their origin in the Shoji-Nishiyama relation, they belong to the class of ordered low-energy coincidence site lattice (CSL) boundaries that can disrupt the connectivity of random HAGBs, resulting in suppressed overall grain boundary degradation upon loading \cite{dong_grain_2021}. In this manner, the presence of such special boundaries with high structural stability can enhance deformation uniformity and, in turn, also the material ductility and strength \cite{dong_grain_2021, guan_pathway_2022}. Even though the different deformation curves (Fig. \ref{def_1100}) of co1100 and co1100-10c can also be affected by the variation in the fcc phase content, in the case of co1100-10 and co1100-20 this effect can be ascribed primarily to the effect of special \tld{}\ang{71} HAGBs. While the rudimentary microstructural observations suggested the stabilization of transformation, the evolution of \tld{}\ang{71} boundaries and their continuing presence in the co1100-20c material indicate that the transformation is not yet fully reversible and in equilibrium even after 20 thermal cycles. Further quantitative characterization is beyond the scope of the present study; nonetheless, it can be concluded that when dealing with such transformation-driven microstructures, advanced microstructural analyses are necessary to justify dissimilarities in the mechanical (and other) properties of the material. From the technological point of view, these non-trivial effects provide compelling means of material performance optimization in relation to carefully selected thermal treatments. 

\section{Conclusions}

Pure polycrystalline cobalt was prepared to have varying microstructural features -- especially the grain size and the content of the residual fcc phase -- using different thermal treatment and cycling around the hcp$\leftrightarrow$fcc transformation temperature. Deformation dynamics were studied in compression complemented by in-situ AE recording and ex-situ SEM/EBSD observations. The EBSD data acquired during interrupted loading were further used to determine the evolution of twin and fcc phase fractions, as well as the types of grain boundaries. Such a combination of methods was selected with the aim of understanding the operation of elementary mechanisms that accommodate plastic deformation in these samples with non-trivial microstructures. The conclusions drawn from this study are as follows:

\begin{itemize}
    \item The observed interconnection of deformation behavior and grain size in cobalt agrees with what was reported for hcp and other metals only to some extent. Even though larger grain size can lead to diminishing flow stress and greater deformability, the presence of fcc grains and special \tld{}\ang{71} misorientation grain boundaries between the hcp martensite variants considerably affect the deformation. The stress-induced fcc$\rightarrow$hcp transformation and the wealth of slip systems in the fcc structure, compared to the hcp structure, allow the fcc grains to accommodate a good amount of strain, improving the ductility and also enhancing the material hardening. On the other hand, the ordered \tld{}\ang{71} grain boundaries act as microstructure stabilizer by disrupting the networks of standard HAGBs and contribute to the deformation uniformity.
    \item Accordingly, the fcc$\rightarrow$hcp transformation dynamics during thermal cycling and their effect on the hcp phase substructure must always be considered. Cycling stabilizes the hcp$\leftrightarrow$fcc transformation rather gradually and even if practically no fcc phase remains in the material upon cooling, material microstructure evolves with further cycling. The presence of special \tld{}\ang{71} boundaries within the hcp phase observed in this work even after 20 thermal cycles confirms that ``perfect'' transformation, i.e. one fcc grain transforming into one hcp grain and vice versa, might be achieved only after much higher number of cycles.  
    \item Martensitic transformations are generally considered a rather powerful source of AE. On the contrary, the stress-induced fcc$\rightarrow$hcp transformation documented in this study takes place rather gradually during loading and does not give rise to detectable AE signals. Moreover, a significant fcc phase fraction is retained in the material even after straining up to \SI{30}{\percent}. Hence, the martensitic transformation in cobalt appears quite sluggish, unlike in many other classes of materials exhibiting this type of load-driven transformation. 
    \item The most pronounced source of AE is the tested samples is presumably the nucleation of mechanical twins in the vicinity of the yield point. The intensity of the signals is greater in the samples subjected to higher annealing temperatures and is further enhanced by thermal cycling. The effect is related to the microstructure ``clean-up'' resulting in the increased mean free path of defect motion. 
    \item The presence of residual fcc phase and special \tld{}\ang{71} boundaries in cobalt makes it an intriguing research subject. Based on the revealed insights into cobalt plasticity, novel research procedures employing phase content and grain boundary engineering might be designed in order to further understand and optimize microstructure and performance of cobalt, with the aim of fulfilling demanding criteria related to contemporary applications.
\end{itemize}

\section*{Acknowledgements}
This work received funding from the Czech Science Foundation (project No. 23-05662S). Financial support by the Operational Programme Johannes Amos Comenius of the MEYS of the Czech Republic, within the frame of the project Ferroic Multifunctionalities (FerrMion) [project No. CZ.02.01.01/00/22\_008/0004591], co-funded by the European Union is also gratefully acknowledged. A.G. would like to thank for funding from the Charles University Grant Agency (Project No. 360721). The authors would like to acknowledge helpful discussions with Prof. Kristián Máthis (Charles University, Czech Republic).

\printcredits

\section*{Data availability}

The data that support the findings of this study are openly available in the Zenodo repository at \newline\noindent \href{https://doi.org/10.5281/zenodo.12759010}{https://doi.org/10.5281/zenodo.12759010} under the \href{https://creativecommons.org/licenses/by-nc/4.0/}{CC-BY 4.0 licence}.




\bibliographystyle{elsarticle-num.bst}



\bibliography{cas-refs}

\begin{thebibliography}{10}
\expandafter\ifx\csname url\endcsname\relax
  \def\url#1{\texttt{#1}}\fi
\expandafter\ifx\csname urlprefix\endcsname\relax\def\urlprefix{URL }\fi
\expandafter\ifx\csname href\endcsname\relax
  \def\href#1#2{#2} \def\path#1{#1}\fi

\bibitem{betteridge_properties_1980}
W.~Betteridge, The properties of metallic cobalt, Progress in Materials Science 24 (1980) 51--142.
\newblock \href {https://doi.org/10.1016/0079-6425(79)90004-5} {\path{doi:10.1016/0079-6425(79)90004-5}}.

\bibitem{sargent_deformation_1983}
P.~M. Sargent, G.~Malakondaiah, M.~F. Ashby, A deformation map for cobalt, Scripta Metallurgica 17~(5) (1983) 625--629.
\newblock \href {https://doi.org/10.1016/0036-9748(83)90390-3} {\path{doi:10.1016/0036-9748(83)90390-3}}.

\bibitem{zheng_grain-size_2007}
G.~Zheng, Grain-size effect on plastic flow in nanocrystalline cobalt by atomistic simulation, Acta Materialia 55~(1) (2007) 149--159.
\newblock \href {https://doi.org/10.1016/j.actamat.2006.07.034} {\path{doi:10.1016/j.actamat.2006.07.034}}.

\bibitem{crone_brief_2008}
W.~C. Crone, A {Brief} {Introduction} to {MEMS} and {NEMS}, in: W.~N. Sharpe (Ed.), Springer {Handbook} of {Experimental} {Solid} {Mechanics}, Springer US, Boston, MA, 2008, pp. 203--228.
\newblock \href {https://doi.org/10.1007/978-0-387-30877-7_9} {\path{doi:10.1007/978-0-387-30877-7_9}}.

\bibitem{edwards_xray_1943}
O.~S. Edwards, H.~Lipson, An {X-ray} study of the transformation of cobalt, Journal of the Institute of Metals 96 (1943) 177--187.

\bibitem{troiano_transformation_1948}
A.~R. Troiano, J.~L. Tokich, The transformation of cobalt, Transactions of the American Institute of Mining and Metallurgical Engineers 175 (1948) 728--741.

\bibitem{nishiyama_crystallography_1978}
Z.~Nishiyama, Crystallography of martensite (general), in: E.~F. Morris, M.~Meshii, C.~M. Wayman, Z.~Nishiyama (Eds.), Martensitic Transformation, Academic Press, 1978, pp. 14--134.
\newblock \href {https://doi.org/10.1016/B978-0-12-519850-9.50007-7} {\path{doi:10.1016/B978-0-12-519850-9.50007-7}}.

\bibitem{laughlin_physical_2014}
D.~Laughlin, K.~Hono, Physical {Metallurgy}, Elsevier, 2014.

\bibitem{schryvers_electron_1997}
D.~Schryvers, Electron {Microscopy} {Studies} of {Martensite} {Microstructures}, Le Journal de Physique IV 07~(C5) (1997) 109--118.
\newblock \href {https://doi.org/10.1051/jp4:1997517} {\path{doi:10.1051/jp4:1997517}}.

\bibitem{xie_transmission_2005}
G.~Xie, M.~Song, K.~Mitsuishi, K.~Furuya, Transmission {Electron} {Microscopy} of {Martensitic} {Transformation} in {Xe}-implanted {Austenitic} 304 {Stainless} {Steel}, Journal of Materials Research 20~(7) (2005) 1751--1757.
\newblock \href {https://doi.org/10.1557/JMR.2005.0218} {\path{doi:10.1557/JMR.2005.0218}}.

\bibitem{chang_high-temperature_2019}
S.-H. Chang, P.-T. Lin, C.-W. Tsai, High-temperature martensitic transformation of {CuNiHfTiZr} high- entropy alloys, Scientific Reports 9~(1) (2019) 19598.
\newblock \href {https://doi.org/10.1038/s41598-019-55762-y} {\path{doi:10.1038/s41598-019-55762-y}}.

\bibitem{bauer_kinetics_2011}
R.~Bauer, E.~A. Jägle, W.~Baumann, E.~J. Mittemeijer, Kinetics of the allotropic hcp–fcc phase transformation in cobalt, Philosophical Magazine 91~(3) (2011) 437--457.
\newblock \href {https://doi.org/10.1080/14786435.2010.525541} {\path{doi:10.1080/14786435.2010.525541}}.

\bibitem{song_recrystallization_2022}
K.~Song, Z.~Li, M.~Fang, Z.~Xiao, Y.~Zhu, Q.~Lei, Recrystallization behavior and phase transformation in a hot-rolled pure cobalt during annealing at the elevated temperature, Materials Science and Engineering: A 845 (2022) 143178.
\newblock \href {https://doi.org/10.1016/j.msea.2022.143178} {\path{doi:10.1016/j.msea.2022.143178}}.

\bibitem{toledano_theory_2001}
P.~Tolédano, G.~Krexner, M.~Prem, H.-P. Weber, V.~Dmitriev, Theory of the martensitic transformation in cobalt, Physical Review B 64~(14) (2001) 144104.
\newblock \href {https://doi.org/10.1103/PhysRevB.64.144104} {\path{doi:10.1103/PhysRevB.64.144104}}.

\bibitem{munier_evolution_1990}
A.~Munier, J.~E. Bidaux, R.~Schaller, C.~Esnouf, Evolution of the microstructure of cobalt during diffusionless transformation cycles, Journal of Materials Research 5~(4) (1990) 769--775.
\newblock \href {https://doi.org/10.1557/JMR.1990.0769} {\path{doi:10.1557/JMR.1990.0769}}.

\bibitem{bidaux_study_1987}
J.-E. Bidaux, R.~Schaller, W.~Benoit, {Study} {of} {the} hcp-fcc {phase} {transition} {in} {cobalt} {by} {internal} {friction} {and} {elastic} {modulus} {measurements} {in} {the} {kHz} {frequency} {range}, Le Journal de Physique Colloques 48~(C8) (1987) C8--477--C8--482.
\newblock \href {https://doi.org/10.1051/jphyscol:1987874} {\path{doi:10.1051/jphyscol:1987874}}.

\bibitem{kuang_latent_2000}
Z.~Q. Kuang, J.~X. Zhang, X.~H. Zhang, K.~F. Liang, P.~C.~W. Fung, Latent heat in the thermoelastic martensitic transformation of {Co}, Scripta Materialia 42~(8) (2000) 795--799.
\newblock \href {https://doi.org/10.1016/S1359-6462(00)00297-9} {\path{doi:10.1016/S1359-6462(00)00297-9}}.

\bibitem{mises_mechanik_1928}
R.~G. v.~Mises, Mechanik der plastischen {Formänderung} von {Kristallen}, ZAMM - Zeitschrift für Angewandte Mathematik und Mechanik 8~(3) (1928) 161--185.
\newblock \href {https://doi.org/10.1002/zamm.19280080302} {\path{doi:10.1002/zamm.19280080302}}.

\bibitem{sanderson_deformation_1972}
C.~C. Sanderson, {Deformation} {of} {polycrystalline} {cobalt}, {PhD}. thesis, University of British Columbia, Vancouver, Canada (1972).

\bibitem{dubos_temperature_2020}
P.-A. Dubos, J.~Fajoui, N.~Iskounen, M.~Coret, S.~Kabra, J.~Kelleher, B.~Girault, D.~Gloaguen, Temperature effect on strain-induced phase transformation of cobalt, Materials Letters 281 (2020) 128812.
\newblock \href {https://doi.org/10.1016/j.matlet.2020.128812} {\path{doi:10.1016/j.matlet.2020.128812}}.

\bibitem{holt_high_1968}
R.~T. Holt, {The} {high} {temperature} {deformation} {of} {cobalt} {single} {crystals}, {PhD}. thesis, The University of British Columbia, Vancouver, Canada (1968).

\bibitem{holt_influence_1972}
R.~T. Holt, E.~Teghtsoonian, The influence of the allotropic transformation on the deformation behavior of pure cobalt, Metallurgical Transactions 3~(9) (1972) 2443--2447.
\newblock \href {https://doi.org/10.1007/BF02647047} {\path{doi:10.1007/BF02647047}}.

\bibitem{korner_weak-beam_1983}
A.~Korner, H.~P. Karnthaler, Weak-beam study of glide dislocations in h.c.p. cobalt, Philosophical Magazine A 48~(3) (1983) 469--477.
\newblock \href {https://doi.org/10.1080/01418618308234904} {\path{doi:10.1080/01418618308234904}}.

\bibitem{marx_strain-induced_2016}
V.~M. Marx, C.~Kirchlechner, B.~Breitbach, M.~J. Cordill, D.~M. Többens, T.~Waitz, G.~Dehm, Strain-induced phase transformation of a thin {Co} film on flexible substrates, Acta Materialia 121 (2016) 227--233.
\newblock \href {https://doi.org/10.1016/j.actamat.2016.09.015} {\path{doi:10.1016/j.actamat.2016.09.015}}.

\bibitem{sort_microstructural_2003}
J.~Sort, J.~Nogués, S.~Suriñach, M.~D. Baró, Microstructural aspects of the hcp-fcc allotropic phase transformation induced in cobalt by ball milling, Philosophical Magazine 83~(4) (2003) 439--455.
\newblock \href {https://doi.org/10.1080/0141861021000047159} {\path{doi:10.1080/0141861021000047159}}.

\bibitem{edalati_high-pressure_2013}
K.~Edalati, S.~Toh, M.~Arita, M.~Watanabe, Z.~Horita, High-pressure torsion of pure cobalt: hcp-fcc phase transformations and twinning during severe plastic deformation, Applied Physics Letters 102~(18) (2013) 181902.
\newblock \href {https://doi.org/10.1063/1.4804273} {\path{doi:10.1063/1.4804273}}.

\bibitem{barry_microstructure_2014}
A.~H. Barry, G.~Dirras, F.~Schoenstein, F.~Tétard, N.~Jouini, Microstructure and mechanical properties of bulk highly faulted fcc/hcp nanostructured cobalt microstructures, Materials Characterization 91 (2014) 26--33.
\newblock \href {https://doi.org/10.1016/j.matchar.2014.02.004} {\path{doi:10.1016/j.matchar.2014.02.004}}.

\bibitem{kapoor_aspects_2009}
R.~Kapoor, B.~Paul, S.~Raveendra, I.~Samajdar, J.~Chakravartty, Aspects of {Dynamic} {Recrystallization} in {Cobalt} at {High} {Temperatures}, Metallurgical and Materials Transactions A 40~(4) (2009) 818--827.
\newblock \href {https://doi.org/10.1007/s11661-009-9782-8} {\path{doi:10.1007/s11661-009-9782-8}}.

\bibitem{seeger_plastische_1963}
A.~Seeger, H.~Kronmüller, O.~Boser, M.~Rapp, Plastische {Verformung} von {Kobalteinkristallen}, physica status solidi (b) 3~(6) (1963) 1107--1125.
\newblock \href {https://doi.org/10.1002/pssb.19630030617} {\path{doi:10.1002/pssb.19630030617}}.

\bibitem{zhang_deformation_2010}
X.~Y. Zhang, Y.~T. Zhu, Q.~Liu, Deformation twinning in polycrystalline {Co} during room temperature dynamic plastic deformation, Scripta Materialia 63~(4) (2010) 387--390.
\newblock \href {https://doi.org/10.1016/j.scriptamat.2010.04.031} {\path{doi:10.1016/j.scriptamat.2010.04.031}}.

\bibitem{fleurier_size_2015}
G.~Fleurier, E.~Hug, M.~Martinez, P.-A. Dubos, C.~Keller, Size effects and {Hall}–{Petch} relation in polycrystalline cobalt, Philosophical Magazine Letters 95~(2) (2015) 122--130.
\newblock \href {https://doi.org/10.1080/09500839.2015.1020351} {\path{doi:10.1080/09500839.2015.1020351}}.

\bibitem{martinez_tem_2017}
M.~Martinez, G.~Fleurier, F.~Chmelík, M.~Knapek, B.~Viguier, E.~Hug, {TEM} analysis of the deformation microstructure of polycrystalline cobalt plastically strained in tension, Materials Characterization 134 (2017) 76--83.
\newblock \href {https://doi.org/10.1016/j.matchar.2017.09.038} {\path{doi:10.1016/j.matchar.2017.09.038}}.

\bibitem{knapek_effect_2020}
M.~Knapek, P.~Minárik, P.~Dobroň, J.~Šmilauerová, M.~M. Celis, E.~Hug, F.~Chmelík, The effect of different thermal treatment on the allotropic fcc{$\leftrightarrow$}hcp transformation and compression behavior of polycrystalline cobalt, Materials 13~(24) (2020) 5775.
\newblock \href {https://doi.org/10.3390/ma13245775} {\path{doi:10.3390/ma13245775}}.

\bibitem{dobron_grain_2011}
P.~Dobroň, F.~Chmelík, S.~Yi, K.~Parfenenko, D.~Letzig, J.~Bohlen, Grain size effects on deformation twinning in an extruded magnesium alloy tested in compression, Scripta Materialia 65~(5) (2011) 424--427.
\newblock \href {https://doi.org/10.1016/j.scriptamat.2011.05.027} {\path{doi:10.1016/j.scriptamat.2011.05.027}}.

\bibitem{lejcek_grain_2010}
P.~Lejček, Grain Boundary Segregation in Metals, Vol. 136 of Springer Series in Material Science, Springer Berlin, 2010.
\newblock \href {https://doi.org/10.1007/978-3-642-12505-8} {\path{doi:10.1007/978-3-642-12505-8}}.

\bibitem{gres_experimental_2023}
A.~Greš, Experimental study of the deformation mechanisms in cobalt by the advanced in-situ techniques, Diploma {thesis}, Charles University, Prague (2023).

\bibitem{hug_size_2019}
E.~Hug, C.~Keller, Size effects and magnetoelastic couplings: a link between hall–petch behaviour and coercive field in soft ferromagnetic metals, Philosophical Magazine 99~(11) (2019) 1297--1326.
\newblock \href {https://doi.org/10.1080/14786435.2019.1580397} {\path{doi:10.1080/14786435.2019.1580397}}.

\bibitem{dieter_mechanical_1988}
G.~E. Dieter, Mechanical Metallurgy, Materials Science and Engineering Series, McGraw-Hill Book Company, 1988.

\bibitem{chmelik_exploring_2012}
K.~Máthis, F.~Chmelík, Exploring plastic deformation of metallic materials by the acoustic emission technique, in: W.~Sikorski (Ed.), Acoustic Emission, IntechOpen, Rijeka, 2012.
\newblock \href {https://doi.org/10.5772/31660} {\path{doi:10.5772/31660}}.

\bibitem{maas_micro-plasticity_2018}
R.~Maaß, P.~M. Derlet, Micro-plasticity and recent insights from intermittent and small-scale plasticity, Acta Materialia 143 (2018) 338--363.
\newblock \href {https://doi.org/10.1016/j.actamat.2017.06.023} {\path{doi:10.1016/j.actamat.2017.06.023}}.

\bibitem{voronenko_acoustic_1982}
B.~I. Voronenko, Acoustic emission during phase transformations in alloys, Metal Science and Heat Treatment 24~(8) (1982) 545--553.
\newblock \href {https://doi.org/10.1007/BF00769364} {\path{doi:10.1007/BF00769364}}.

\bibitem{manosa_acoustic_1990}
L.~Mañosa, A.~Planes, D.~Rouby, J.~L. Macqueron, Acoustic emission in martensitic transformations, Acta Metallurgica et Materialia 38~(9) (1990) 1635--1642.
\newblock \href {https://doi.org/10.1016/0956-7151(90)90006-3} {\path{doi:10.1016/0956-7151(90)90006-3}}.

\bibitem{toth_acoustic_2020}
L.~Z. Tóth, L.~Daróczi, E.~Panchenko, Y.~Chumlyakov, D.~L. Beke, Acoustic {Emission} {Characteristics} and {Change} the {Transformation} {Entropy} after {Stress}-{Induced} {Martensite} {Stabilization} in {Shape} {Memory} {Ni53Mn25Ga22} {Single} {Crystal}, Materials 13~(9) (2020) 2174.
\newblock \href {https://doi.org/10.3390/ma13092174} {\path{doi:10.3390/ma13092174}}.

\bibitem{bohlen_effect_2006}
J.~Bohlen, P.~Dobroň, E.~Meza~Garcia, F.~Chmelík, P.~Lukáč, D.~Letzig, K.~U. Kainer, The effect of grain size on the deformation behaviour of magnesium alloys investigated by the acoustic emission technique, Advanced Engineering Materials 8~(5) (2006) 422--427.
\newblock \href {https://doi.org/10.1002/adem.200600023} {\path{doi:10.1002/adem.200600023}}.

\bibitem{wang_deformation_2014}
X.~Wang, L.~Jiang, A.~Luo, J.~Song, Z.~Liu, F.~Yin, Q.~Han, S.~Yue, J.~Jonas, Deformation of twins in a magnesium alloy under tension at room temperature, Journal of Alloys and Compounds 594 (2014) 44--47.
\newblock \href {https://doi.org/10.1016/j.jallcom.2014.01.100} {\path{doi:10.1016/j.jallcom.2014.01.100}}.

\bibitem{krajnak_influence_2019}
T.~Krajňák, P.~Minárik, J.~Stráská, J.~Gubicza, K.~Máthis, M.~Janeček, Influence of the initial state on the microstructure and mechanical properties of {AX41} alloy processed by {ECAP}, Journal of Materials Science 54~(4) (2019) 3469--3484.
\newblock \href {https://doi.org/10.1007/s10853-018-3033-6} {\path{doi:10.1007/s10853-018-3033-6}}.

\bibitem{drozdenko_investigating_2016}
D.~Drozdenko, J.~Bohlen, S.~Yi, P.~Minárik, F.~Chmelík, P.~Dobroň, Investigating a twinning–detwinning process in wrought {Mg} alloys by the acoustic emission technique, Acta Materialia 110 (2016) 103--113.
\newblock \href {https://doi.org/10.1016/j.actamat.2016.03.013} {\path{doi:10.1016/j.actamat.2016.03.013}}.

\bibitem{watanabe_grain_2011}
T.~Watanabe, Grain boundary engineering: historical perspective and future prospects, Journal of Materials Science 46~(12) (2011) 4095--4115.
\newblock \href {https://doi.org/10.1007/s10853-011-5393-z} {\path{doi:10.1007/s10853-011-5393-z}}.

\bibitem{dong_grain_2021}
X.~Dong, N.~Li, Y.~Zhou, H.~Peng, Y.~Qu, Q.~Sun, H.~Shi, R.~Li, S.~Xu, J.~Yan, Grain boundary character and stress corrosion cracking behavior of {Co}-{Cr} alloy fabricated by selective laser melting, Journal of Materials Science \& Technology 93 (2021) 244--253.
\newblock \href {https://doi.org/10.1016/j.jmst.2021.03.063} {\path{doi:10.1016/j.jmst.2021.03.063}}.

\bibitem{guan_pathway_2022}
X.~Guan, Z.~Jia, S.~Liang, F.~Shi, X.~Li, A pathway to improve low-cycle fatigue life of face-centered cubic metals via grain boundary engineering, Journal of Materials Science \& Technology 113 (2022) 82--89.
\newblock \href {https://doi.org/10.1016/j.jmst.2021.09.063} {\path{doi:10.1016/j.jmst.2021.09.063}}.

\end{thebibliography}

\end{document}